\documentclass[pra,twocolumn,showpacs,preprintnumbers,amsmath,amssymb]{revtex4}
\usepackage{bm, psfrag, graphics}
\begin{document}
\title{Time-dependent quantum many-body theory of identical bosons in a double well: Early time ballistic interferences of fragmented and number entangled states}
\author{David J. Masiello}
\email{masiello@chem.northwestern.edu}
\altaffiliation[Present address: ]{Department of Chemistry, Northwestern University, Evanston, IL 60208-3113, USA}
\affiliation{Department of Chemistry, University of Washington, Seattle, Washington 98195-1700, USA}
\author{William P. Reinhardt}
\email{rein@chem.washington.edu}
\affiliation{Department of Chemistry, University of Washington, Seattle, Washington 98195-1700, USA}
\date{February 4, 2007}       
\begin{abstract}
A time-dependent multiconfigurational self-consistent field theory is presented to describe the many-body dynamics of a gas of identical bosonic atoms confined to an external trapping potential at zero temperature from first principles. A set of generalized evolution equations are developed, through the time-dependent variational principle, which account for the complete and self-consistent coupling between the expansion coefficients of each configuration and the underlying one-body wave functions within a restricted two state Fock space basis that includes the full effects of the condensate's mean field as well as atomic correlation. The resulting dynamical equations are a classical Hamiltonian system and, by construction, form a well-defined initial value problem. They are implemented in an efficient numerical algorithm. An example is presented, highlighting the generality of the theory, in which the ballistic expansion of a fragmented condensate ground state is compared to that of a macroscopic quantum superposition state, taken here to be a highly entangled number state, upon releasing the external trapping potential. Strikingly different many-body matter-wave dynamics emerge in each case, accentuating the role of both atomic correlation and mean-field effects in the two condensate states.
\end{abstract}
\pacs{03.75.Kk, 03.75.-b, 05.30.Jp, 03.75.Gg}
\maketitle

\section{Introduction}
Understanding the quantum many-body structural and dynamical properties of the trapped gaseous Bose-Einstein condensate (BEC) lies at the heart of many experimental and theoretical research efforts worldwide. Remarkably sensitive matter-wave interferometry \cite{Shin2004a,Shin2004b,Saba05}, based upon an atomic BEC source, represents just one potentially useful experimental tool of technological import, while the demonstration of coherent macroscopic superposition of millions of Bose-Einstein condensed atoms, may, one day, be turned from dream to reality, possibly answering questions fundamental to the theory of quantum mechanics \cite{Leggett05}. These examples, spanning both practical and deeply fundamental extremes, have either been already realized \cite{Shin2004a,Shin2004b,Saba05} or the first fundamental steps have been achieved \cite{Albiez04} in the context of a BEC made of identical atoms confined to an external double-well trapping potential. In order to complement these experimental accomplishments, theoretical knowledge of the double-well condensate's structure and dynamics, beyond that of simple models, is now in great demand.

On the theoretical front, the time-independent quantum many-body structure of the double-well condensate is beginning to be explored with powerful first principles methods adapted from the quantum-chemical theory of many-electron atomic and molecular systems \cite{Ostlund,Levine} at equilibrium and dissociation, reflecting the strong analogy between a rigorous description of molecular dissociation and that of BEC fragmentation. Such methods involve the simultaneous variational optimization of both the Fock space expansion coefficients and the one-body wave functions (orbitals) underlying each Fock state \cite{Masiello05,Ced06}. These methods build in, both, the full effects of the condensate's mean field and the correlations arising between atoms in different Fock states, and are capable of describing the many-body ground and excited states of the system at any barrier height for either symmetric or asymmetric trapping potentials. The unique role of spatial symmetry breaking in high-lying macroscopic self-trapped and superposition states of the double-well BEC has also recently been explored, for the first time, in detail, at a first principles level \cite{Masiello06}.

Building upon this detailed many-body structure comes the next frontier in the first principles theory of the double-well condensate: the dynamics. It is useful to learn from what has already been accomplished in the explicitly time-dependent many-body approaches to atomic and molecular dynamics. Dynamical methods that are derivable from the principle of least action \cite{Jose,Goldstein}, the time-dependent variational principle \cite{Kramer}, or other equivalent variational approaches that provide a representation of the time-dependent Schr\"odinger equation \cite{Broeckhove88} are regarded, by the authors, to be especially inspiring as they enjoy, by construction, various symmetries and associated conservation laws, and lead to well-defined evolution equations that include the complete coupling of the dynamical variables within a chosen parameter space. In particular, description of the reactive chemical dynamics of general atomic or molecular systems, beyond the adiabatic and Born-Oppenheimer approximations, by a single determinant of coupled electronic parameters and nuclear coordinates and momenta \cite{Deumens94} represents one notable application of the time-dependent variational principle.  Generalization of this work to the case of multiple determinants \cite{Weiner91} represents another. Also, recently, formally similar well-defined equations of motion describing the dynamics of the classical electromagnetic field coupled completely to its atomic sources of charge and current \cite{Masiello04,Masiello05a} have been developed with the aid of the time-dependent variational calculus. The Dirac-Frenkel variational principle \cite{Dirac30,Frenkel} applied to a multiconfigurational state vector for distinguishable bosons has been successful in characterizing, {\it inter alia}, the vibrational dynamics occurring in general polyatomic molecules \cite{Meyer00}. This same approach has been used to study the transition of a small and fixed number of indistinguishable bosons from a coherent to a fragmented and finally to a fermionized ground state by imaginary time propagation \cite{Schmelcher06}. As noted in Ref. \cite{Broeckhove88}, the Dirac-Frenkel form of the variational principle and the time-dependent variational principle used here in Sec. III. may yield superficially different looking evolution equations; however, if the variations are carried out fully and correctly, they will both give identical results. A notable time-dependent treatment of the double-well BEC, already in the literature, that intuitively describes the dynamics of a double-well BEC by an improved two-mode model with time-dependent Fock space expansion coefficients but static Gross-Pitaevskii orbitals underlying each Fock state may be found in Ref. \cite{Berg06}.

In the spirit of the above approaches to dynamics, a time-dependent multiconfigurational bosonic self-consistent field state vector is presented in Sec. II. that is capable of describing the many-body dynamics of a gas of identical bosonic atoms at zero temperature confined to an external trapping potential that can be continuously deformed from a single well to a double well. It is parametrized by complex-valued time-dependent Fock state expansion coefficients and atomic orbitals, which inherit the role of generalized coordinates in a nonlinear phase space. Equations of motion associated with this parametrized state vector are derived in Sec. III. by application of the time-dependent variational principle to the many-body Schr\"odinger Lagrangian. Properties of the evolution equations are discussed and a symplectic transformation is made to map the phase space coordinates to a real form that is amenable to numerical implementation.  Lastly, in Sec. IV., numerical examples are presented in which the ballistic expansion dynamics of a fragmented ground state is compared to that of a macroscopic quantum superposition state (also called a Schr\"odinger cat state) of the BEC. These results are contrasted with a time-dependent Gross-Pitaevskii approximation of ballistic BEC expansion.

\section{Multiconfigurational state vector ansatz and overlap}
The most general time-dependent state vector representing a gas of $N$ identical bosonic atoms confined to an external double-well trapping potential at zero temperature should be flexible enough to describe the motion of atoms, in all possible arrangements, between the two trap minima, and, in addition, the dynamics of the state associated with each particular arrangement. Here it is assumed that two wells implies two Fock states. As the double-well experiments that are of primary interest to us involve symmetric trapping potentials \cite{Shin2004a,Shin2004b,Saba05,Albiez04} we further specialize to this symmetry \cite{fn13}; therefore, within this restricted model space, a time-dependent state is sought that accounts for all possible arrangements of atoms. The collection of Fock states 
\begin{equation}
|\Phi_{N_1}[{\bm\phi}]\rangle\equiv|N_1,N_2\rangle=(\hat b_1^\dagger)^{N_1}(\hat b_2^\dagger)^{N_2}|{\textrm{vac}}\rangle/\sqrt{N_1!N_2!},
\end{equation}
for $N_1=0,\ldots,N$ and $N_2=N-N_1,$ with $N_k$ atoms restricted to two macroscopically occupied states, $k=1,2,$ with the spatially orthogonal symmetric and antisymmetric one-body orbitals $\phi_k({\bf x},t)$ underlying each Fock state, provides such a basis. Therefore, in this restricted space, the most general completely symmetric many-body state may be written as the superposition
\begin{equation}
\label{ansatz}
\begin{split}
|{\bf d}(t);{\bm\phi}(t)\rangle&=\sum_{N_1=0}^N|\Phi_{N_1}[{\bm\phi}(t)]\rangle d_{N_1}(t),
\end{split}
\end{equation}
where the vector ${\bf d}\equiv{\bf d}(t)=[d_0(t),\ldots,d_N(t)]$ and ${\bm\phi}\equiv{\bm\phi}(t)=[\phi_1({\bf x},t),\phi_2({\bf x},t)]$ are the collection of all $N+1$ time-dependent expansion coefficients and symmetric and antisymmetric space- and time-dependent orbitals. Each term in the series, representing a particular arrangement of atoms between the two wells, is a called a configuration. This ansatz, which is a superposition of configurations, is a time-dependent generalization of the multiconfigurational state promoted in Refs. \cite{Masiello05,Ced06}, where both the linear Fock state expansion coefficients and nonlinear symmetrized product of orbitals within $|\Phi_{N_1}[{\bm\phi}]\rangle$ are variationally optimized and mutually self-consistent; it is a time-dependent multiconfigurational bosonic self-consistent field (TDMCBSCF) state vector. Both the expansion coefficients ${\bf d}$ and the orbitals ${\bm\phi}$ of each configuration $|\Phi_{N_1}[{\bm\phi}]\rangle$ are complex-valued and time dependent \cite{fn8}. It will be demonstrated that these parameters form a set of generalized coordinates whose canonically conjugate momenta are proportional to their respective complex conjugates. Such a parametrized TDMCBSCF state spans a sufficiently general phase space to allow, upon application of the variational calculus, for a set of well-defined first order coupled nonlinear evolution equations with solutions enjoying a rich dynamics influenced by the full coupling between condensate's mean field and the quantum-mechanical correlations arising between atoms in different configurations.

A basic ingredient in the variational approach to deriving dynamical equations is the overlap of two many-body states (\ref{ansatz}). This multiconfigurational overlap involves the overlap of two single-configurational states $|\Phi_{N_1}[{\bm\phi}]\rangle.$ Projection of $|\Phi_{N_1}[{\bm\phi}]\rangle$ onto the coordinate basis yields the single-permanental wave function
\begin{equation}
\label{perm}
\begin{split}
\Phi_{N_1}&(1,\ldots,N)=\langle 1,\ldots,N|\Phi_{N_1}[{\bm\phi}]\rangle\\
&={\cal S}\{\phi_1(1)\cdots\phi_1(N_1)\phi_2(N_1+1)\cdots\phi_2(N)\},
\end{split}
\end{equation}
where, for simplicity, the variable $j\equiv({\bf x}_j,t)$ for $j=1,\ldots,N$ and ${\cal S}=(\sqrt{N!N_1!N_2!})^{-1}\sum_PP$ is the symmetrizer with permutation operator $P.$ The spatial overlap of two such single-permanental wave functions is expressible in a form that is reminiscent of that for determinants \cite{Lowdin55a}, {\it i.e.},
\begin{equation}
\label{over}
\begin{split}
\langle&\Phi_{N_1}[{\bm\phi}^*]|\Phi_{N_1}[{\bm\phi}]\rangle\\
&={\textstyle\int}\langle\Phi_{N_1}[{\bm\phi}^*]|1,\ldots,N\rangle\langle 1,\ldots,N|\Phi_{N_1}[{\bm\phi}]\rangle d1\cdots dN\\
&={\textstyle\int}{\cal S}\{\phi^{*}_1(1)\cdots\phi^{*}_1(N_1)\phi^{*}_2(N_1+1)\cdots\phi^{*}_2(N)\}\\
&\hspace{0.25cm}\times{\cal S}\{\phi_1(1)\cdots\phi_1(N_1)\phi_2(N_1+1)\cdots\phi_2(N)\}d1\cdots dN\\
&=\sqrt{\frac{N!}{N_1!N_2!}}{\cal S}\{\langle\phi_1|\phi_1\rangle_1\cdots\langle\phi_1|\phi_1\rangle_{N_1}\\
&\hspace{2.25cm}\times\langle\phi_2|\phi_2\rangle_{N_1+1}\cdots\langle\phi_2|\phi_2\rangle_N\}\\
&=(N_1!N_2!)^{-1}{\textrm{perm}}\,\Delta({\bm\phi}^*,{\bm\phi}),
\end{split}
\end{equation}
where a partial integration and ${\cal S}^2=\sqrt{N!/N_1!N_2!}{\cal S}$ has been used, and $dj\equiv d^3x_j.$ The $N\times N$ orbital overlap matrix, denoted by $\Delta,$ reduces to
\begin{equation}
\begin{split}
\Delta({\bm\phi}^*,{\bm\phi})&=
\left[
\begin{array}{ccc}
\Delta_{N_1\times N_1}(\phi^*_1,\phi_1)&\vline&\Delta_{N_1\times N_2}(\phi^*_1,\phi_2)\\
&\vline&\\ 
\hline&\vline&\\
\Delta_{N_2\times N_1}(\phi^*_2,\phi_1)&\vline&\Delta_{N_2\times N_2}(\phi^*_2,\phi_2)\\
\end{array}
\right]\\
&=
\left[
\begin{array}{ccc}
\Delta_{N_1\times N_1}(\phi^*_1,\phi_1)&\vline&0_{N_1\times N_2}\\
&\vline&\\ 
\hline&\vline&\\
0_{N_2\times N_1}&\vline&\Delta_{N_2\times N_2}(\phi^*_2,\phi_2)\\
\end{array}
\right].
\end{split}
\end{equation}
since the orbitals are orthogonal. The permanent of $\Delta$ is $N_1!N_2!,$ which is related to the product of the dimensions of the $N_1\times N_1$ upper left and $N_2\times N_2$ lower right blocks. Hence, the overlap of two permanents is the permanent of the overlaps and $|\Phi_{N_1}[{\bm\phi}]\rangle$ is normalized to unity. Note that time has not been integrated upon in Eq. (\ref{over}) so that the overlap remains time dependent.

All multiconfigurational overlaps involve only single-configurational overlaps from the same configuration as different configurations are automatically orthogonal by symmetry. Together with Eqs. (\ref{ansatz})-(\ref{over}), the full multiconfigurational state overlap is given by
\begin{equation}
\begin{split}
S[{\bf d}^*,{\bm\phi}^*;{\bf d},{\bm\phi}]&=\langle{\bf d};{\bm\phi}|{\bf d};{\bm\phi}\rangle\\
&=\sum_{N_1=0}^N(N_1!N_2!)^{-1}d^*_{N_1}[{\textrm{perm}}\,\Delta({\bm\phi}^*,{\bm\phi})]d_{N_1}.
\end{split}
\end{equation}
Since the Fock state expansion coefficients satisfy $\sum_{N_1=0}^N|d_{N_1}|^2=1,$ the time-dependent multiconfigurational state (\ref{ansatz}) is unit normalized.

In the following, derivatives of the overlap kernel $S$ with respect to $\bm\phi$ and ${\bm\phi}^*$ (and $\bf d$ and ${\bf d}^*$) will be needed to construct dynamical equations.  Employing Laplace's expansion for permanents, {\it i.e.}, ${\textrm{perm}}\,\Delta=\sum_{N_1}\Delta_{N_1N_1'}[{\textrm{minor}}\,\Delta]_{N_1N_1'}$ by expansion along column $N_1',$ these basic derivatives are
\begin{equation}
\begin{split}
\frac{\partial S}{\partial\phi_{k}}&=\sum_{N_1=0}^N(N_1!N_2!)^{-1}d^*_{N_1}\frac{\partial[{\textrm{perm}}\, \Delta]}{\partial\phi_{k}}d_{N_1}\\
&=\sum_{N_1=0}^Nd^*_{N_1}N_k\phi_{k}^*d_{N_1}
\end{split}
\end{equation}
and
\begin{equation}
\begin{split}
\frac{\partial^2S}{\partial\phi_{k}^*\partial\phi_{l}}&=\sum_{N_1=0}^N(N_1!N_2!)^{-1}d^*_{N_1}\frac{\partial^2[{\textrm{perm}}\, \Delta]}{\partial\phi_{k}^*\partial\phi_{l}}d_{N_1}\\
&=\sum_{N_1=0}^Nd^*_{N_1}[N_{l}\delta_{kl}+N_{l}(N_k-\delta_{kl})\phi_{k}\phi^*_{l}]d_{N_1}
\end{split}
\end{equation}
for $k,l=1,2.$ Other derivatives of import, such as $\partial^2S/\partial{\bf d}^*\partial{\bm\phi}$ or $\partial^2S/\partial{\bf d}^*\partial{\bf d},$ can either be obtained from these derivatives by complex conjugation or by differentiation with respect to $\bf d$ and ${\bf d}^*.$ These latter derivatives are simple to compute from $S.$

\section{The time-dependent variational principle}
Upon choosing a functional form for the time-dependent state vector, the time-dependent variational principle (TDVP) \cite{Kramer}, which is derived from the principle of least action applied to the many-body Schr\"odinger Lagrangian, generates a set of first order Hamiltonian equations of motion that form a well-defined initial value problem together with the initial values for the dynamical variables. The choice of parametrized state vector and its numerical representation are the only approximations involved in this approach. Where the numerical representation of the state is systematically improvable, {\it e.g.}, by enriching the computational basis with increased basis functions or grid points, the TDVP generates a hierarchy of systematically improvable equations of motion that form an approximation to the exact dynamical equations. Indeed, in the limit of a complete basis, the TDVP yields, in this case, the exact time-dependent many-body Schr\"odinger equation for $N$ atoms in two macroscopically occupied states. The TDVP equations and their solutions enjoy, by construction, a number of symmetries and their corresponding conservation laws \cite{Kramer}, and, in addition, preserve the orthogonality of the symmetric and antisymmetric orbitals $\phi_1$ and $\phi_2$ without additional equations of constraint \cite{fn13}. Furthermore, the TDVP equations exhibit the complete coupling between the dynamical variables allowed within the chosen state vector ansatz. This latter property, which is not necessarily guaranteed in other time-dependent approaches where intuition guides in the determination of couplings, is automatically achieved through the machinery of the calculus of variations; see, {\it e.g.}, Ref. \cite{Gelfand}.

The many-body Schr\"odinger Lagrangian $L\equiv L[{\bf d}^*,{\bm\phi}^*;{\bf d},{\bm\phi}]$ for a system of $N$ interacting bosonic atoms, taken with respect to the multiconfigurational state (\ref{ansatz}), is
\begin{equation}
\label{L}
\begin{split}
L&=\langle{\bf d};{\bm\phi}|i\hbar(d/dt)-\hat H|{\bf d};{\bm\phi}\rangle\\
&=(i\hbar/2)\langle{\bf d};{\bm\phi}|\stackrel{\longrightarrow}{(d/dt)}|{\bf d};{\bm\phi}\rangle-(i\hbar/2)\langle{\bf d};{\bm\phi}|\stackrel{\longleftarrow}{(d/dt)}|{\bf d};{\bm\phi}\rangle\\
&\hspace{0.5cm}-\langle{\bf d};{\bm\phi}|\hat H|{\bf d};{\bm\phi}\rangle,
\end{split}
\end{equation}
where $E\equiv E[{\bf d}^*,{\bm\phi}^*;{\bf d},{\bm\phi}]=\langle{\bf d};{\bm\phi}|\hat H|{\bf d};{\bm\phi}\rangle$ is the many-body energy and where the pure surface term $-(i\hbar/2)(d/dt)S$ has been added in the second line for symmetrization. Lagrangians differing only by a surface term always yield the same equations of motion and are called {\it equivalent.} Expansion of the ket total time derivative $\stackrel{\longrightarrow}{(d/dt)}=(\partial/\partial\xi)\dot\xi$ and the bra total time derivative $\stackrel{\longleftarrow}{(d/dt)}=(\partial/\partial\xi^*)\dot{\xi}^*,$ with $\xi=[{\bf d},{\bm\phi}],$ results in 
\begin{equation}
\label{L2}
\begin{split}
L&=(i\hbar/2)\Bigl[\frac{\partial S}{\partial{\bf d}}\dot{\bf d}+\frac{\partial S}{\partial{\bm\phi}}\dot{\bm\phi}-\frac{\partial S}{\partial{\bf d}^*}\dot{\bf d}^*-\frac{\partial S}{\partial{\bm\phi}^*}\dot{\bm\phi}^*\Bigr]-E\\
&=(i\hbar/2)\Bigl[\sum_{N_1=0}^N\Bigl\{\frac{\partial S}{\partial d_{N_1}}\dot d_{N_1}-\frac{\partial S}{\partial d^*_{N_1}}\dot d^*_{N_1}\Bigr\}\\
&\hspace{1.25cm}+\sum_{k=1,2}\sum_{q=1}^K\Bigl\{\frac{\partial S}{\partial\phi_{kq}}\dot\phi_{kq}-\frac{\partial S}{\partial \phi^*_{kq}}\dot\phi^*_{kq}\Bigr\}\Bigr]-E,
\end{split}
\end{equation}
where the orbitals $\phi_k$ have been expanded onto an arbitrary real-valued basis $\{g_q({\bf x})\}_{q=1}^K$ of rank $K$ so that $\phi_k({\bf x},t)=\sum_{q=1}^K g_q({\bf x})\phi_{kq}(t).$ In this basis, the complex-valued expansion coefficients $\phi_{kq}\equiv\phi_{kq}(t)$ are explicitly time dependent and inherit the role of dynamical variable. The details of this computational basis will be discussed in greater detail in Sec. IV.

The many-body Hamiltonian, which appears in the above Lagrangian, is given in second quantization by
\begin{equation}
\label{ham}
\begin{split}
\hat H&={\textstyle\int}\hat{\Psi}^\dagger({\bf x},t)h({\bf x})\hat{\Psi}({\bf x},t)d^{3}x\\
&+({1}/{2}){\textstyle\int}\hat{\Psi}^\dagger({\bf x},t)\hat{\Psi}^\dagger({\bf x}',t)V({\bf x},{\bf x}')\hat{\Psi}({\bf x}',t)\hat{\Psi}({\bf x},t)d^{3}xd^{3}x'\\
&=\sum_{kl=1,2}h_{kl}(t)\hat b^\dagger_k\hat b_{l}+(1/2)\sum_{klmn=1,2}V_{klmn}(t)\hat b^\dagger_k\hat b^\dagger_l\hat b_{n}\hat b_{m},
\end{split}
\end{equation}
where the $\hat b_k$ and $\hat b^\dagger_l$ are basic boson annihilation and creation operators making up the boson field operators 
\begin{equation}
\begin{split}
\hat\Psi({\bf x},t)&=\phi_1({\bf x},t)\hat b_1+\phi_2({\bf x},t)\hat b_2\\
\hat\Psi^\dagger({\bf x},t)&=\phi_1^*({\bf x},t)\hat b_1^\dagger+\phi_2^*({\bf x},t)\hat b_2^\dagger,
\end{split}
\end{equation}
and $h_{kl}=\langle\phi_k|h|\phi_{l}\rangle,$ and $V_{klmn}=\langle \phi_k\phi_l|V|\phi_{m}\phi_{n}\rangle$ are matrix elements of the one-body Hamiltonian $h({\bf x})=(-\hbar^2/2m)\nabla^2+V_{\textrm{ext}}({\bf x})$ and two-body atom-atom interaction potential $V({\bf x},{\bf x}')=(4\pi\hbar^2a/m)\delta({\bf x}-{\bf x}')$ in the contact approximation \cite{Huang1987a} with repulsive $s$-wave scattering length $a.$ In terms of this Hamiltonian, the energy is
\begin{equation}
\label{E}
E[{\bf d}^*,{\bm\phi}^*;{\bf d},{\bm\phi}]=\sum_{kl=1,2}h_{kl}\gamma_{kl}+(1/2)\sum_{klmn=1,2}V_{klmn}\Gamma_{klnm},
\end{equation}
where $\gamma_{kl}$ and $\Gamma_{klmn}$ are Fock space matrix elements of the parametrized complex-valued one- and two-body reduced density matrices $\gamma(1,1')$ and $\Gamma(1,1';2,2').$ Their functional form will be specified in Sec. III. C. Note that the many-body energy expectation (\ref{E}), which is a function of the dynamical variables, is multiconfigurational; it is not the energy of a single configuration but, rather, represents the energy of all interacting configurations within the restricted Fock space.

\subsection{The variational calculus}
Carrying out the full calculus of variations or the principle of least action \cite{Goldstein,Jose} on the many-body Schr\"odinger Lagrangian (\ref{L2}) generates a set of well-defined equations of motion that approximate the exact time-dependent many-body Schr\"odinger equation for a system of $N$ identical bosonic atoms at zero temperature in an external trapping potential $V_{\textrm{ext}}.$ Using the collective notation $\xi=[{\bf d},{\bm\phi}]$ for simplicity and generality, these TDVP equations \cite{Goldstein,Jose} are now constructed from the Lagrangian $L=(i\hbar/2)[(\partial S/\partial\xi)\dot\xi-(\partial S/\partial\xi^*)\dot\xi^*]-E$ in Eq. (\ref{L2}).

The momentum conjugate to $\xi$ is 
\begin{equation}
\frac{\partial L}{\partial\dot\xi}=(i\hbar/2)\frac{\partial S}{\partial\xi}
\end{equation}
so that its total time derivative becomes
\begin{equation}
\begin{split}
(d/dt)\frac{\partial L}{\partial\dot\xi}&=(i\hbar/2)\Bigl[\frac{\partial}{\partial\xi}\dot\xi+\frac{\partial}{\partial\xi^*}\dot\xi^*\Bigr]\frac{\partial S}{\partial\xi}\\
&=(i\hbar/2)\Bigl[\frac{\partial^2 S}{\partial\xi\partial\xi}\dot\xi+\frac{\partial^2 S}{\partial\xi^*\partial\xi}\dot\xi^*\Bigr].
\end{split}
\end{equation}
Lastly, the derivative of $L$ with respect to $\xi$ itself is
\begin{equation}
\frac{\partial L}{\partial\xi}=(i\hbar/2)\Bigl[\frac{\partial^2 S}{\partial\xi\partial\xi}\dot\xi-\frac{\partial^2 S}{\partial\xi\partial\xi^*}\dot\xi^*\Bigr]-\frac{\partial E}{\partial\xi}.
\end{equation}
The TDVP equations, which are of first order in time, may then be built up from $(d/dt)\partial L/\partial\dot\xi=\partial L/\partial\xi$ and its complex conjugate, {\it i.e.},
\begin{equation}
\label{tdvp}
-i\hbar\frac{\partial^2 S}{\partial\xi\partial\xi^*}\dot\xi^*=\frac{\partial E}{\partial\xi}\hspace{.5cm}\textrm{and}\hspace{.5cm}i\hbar\frac{\partial^2 S}{\partial\xi^*\partial\xi}\dot\xi=\frac{\partial E}{\partial\xi^*},
\end{equation}
where it has been assumed that the mixed partial derivatives of $S$ commute with respect to $\xi$ and $\xi^*.$ It is these derivatives of the overlap that were computed, in part, in Sec. II. Eqs. (\ref{tdvp}) form a classical Hamiltonian system.

Having given a general derivation of the Hamilton's (TDVP) equations for the collective dynamical variables $\xi$ and $\xi^*,$ the complete set of TDMCBSCF dynamical equations are now presented. Following the same machinery as previously described, they are:
\begin{widetext}
\begin{subequations}
\label{eom1}
\begin{eqnarray}
-i\hbar\Bigl[\sum_{N_1'=0}^N\frac{\partial^2 S}{\partial d_{N_1}\partial d^*_{N_1'}}\dot d^*_{N_1'}+\sum_{l=1,2}\sum_{q'=1}^K\frac{\partial^2 S}{\partial d_{N_1}\partial\phi^*_{lq'}}\dot\phi^*_{lq'}\Bigr]=\frac{\partial E}{\partial d_{N_1}}\\
i\hbar\Bigl[\sum_{N_1'=0}^N\frac{\partial^2 S}{\partial d^*_{N_1}\partial d_{N_1'}}\dot d_{N_1'}+\sum_{l=1,2}\sum_{q'=1}^K\frac{\partial^2 S}{\partial d^*_{N_1}\partial\phi_{lq'}}\dot\phi_{lq'}\Bigr]=\frac{\partial E}{\partial d^*_{N_1}}\\
-i\hbar\Bigl[\sum_{N_1'=0}^N\frac{\partial^2 S}{\partial\phi_{kq}\partial d^*_{N_1'}}\dot d^*_{N_1'}+\sum_{l=1,2}\sum_{q'=1}^K\frac{\partial^2 S}{\partial\phi_{kq}\partial\phi^*_{lq'}}\dot\phi^*_{lq'}\Bigr]=\frac{\partial E}{\partial\phi_{kq}}\\
i\hbar\Bigl[\sum_{N_1'=0}^N\frac{\partial^2 S}{\partial\phi^*_{kq}\partial d_{N_1'}}\dot d_{N_1'}+\sum_{l=1,2}\sum_{q'=1}^K\frac{\partial^2 S}{\partial\phi^*_{kq}\partial\phi_{lq'}}\dot\phi_{lq'}\Bigr]=\frac{\partial E}{\partial\phi^*_{kq}}
\end{eqnarray}
\end{subequations}
\end{widetext}
These equations are closed and coupled, as is already evident, and will be shown to be highly nonlinear. Once ${\bf d},\, {\bf d}^*,\, {\bm\phi},$ and ${\bm\phi}^*$ are specified at some point in time, then Eqs. (\ref{eom1}) form a well-defined initial value problem. The resulting dynamics unfolds in a generalized phase space of dimension $2(N+1)+4K$ whose coordinates are the generalized positions and momenta ${\bf d},\, {\bf d}^*,\, {\bm\phi},$ and ${\bm\phi}^*.$

By restricting the time-dependent state vector (\ref{ansatz}) to a single configuration, the TDVP would generate a set time-dependent SCF or Hartree-Fock equations for identical bosons in two orbitals.  Such a derivation has already been performed in the identical fermionic case for arbitrarily many orbitals \cite{Kermin76,Ring80}. An analogous set of time-dependent mean-field equations has also recently been derived for identical bosons in arbitrarily many orbitals \cite{Ced06a}.

\subsection{Equations of motion in canonical form}
The TDVP equations of motion presented in Eqs. (\ref{eom1}), which contain the full coupling allowed within the ansatz (\ref{ansatz}) between the Fock state expansion coefficients $\bf d$ of each configuration $|\Phi_{N_1}[{\bm\phi}]\rangle$ and the underlying orbitals ${\bm\phi},$ may be collected into matrix form to simplify and, simultaneously, highlight their structure. Introducing the notation
\begin{equation}
C_{\xi\xi}=\frac{\partial^2 S}{\partial{\xi^*}\partial\xi},
\end{equation}
Eqs. (\ref{eom1}) become:
\begin{subequations}
\label{eom2}
\begin{eqnarray}
-i\hbar C^*_{{\bf dd}}\dot{\bf d}^*-i\hbar C^*_{{\bf d}{\bm\phi}}\dot{\bm\phi}^*&=&{\partial E}/{\partial{\bf d}}\\
i\hbar C_{\bf dd}\dot{\bf d}+i\hbar C_{{\bf d}{\bm\phi}}\dot{\bm\phi}&=&{\partial E}/{\partial{\bf d}^*}\\
-i\hbar C^T_{{\bf d}{\bm\phi}}\dot{\bf d}^*-i\hbar C^*_{{\bm\phi}{\bm\phi}}\dot{\bm\phi}^*&=&{\partial E}/{\partial{\bm\phi}}\\
i\hbar C^\dagger_{{\bf d}{\bm\phi}}\dot{\bf d}+i\hbar C_{{\bm\phi}{\bm\phi}}\dot{\bm\phi}&=&{\partial E}/{\partial{\bm\phi}^*}
\end{eqnarray}
\end{subequations}
These TDMCBSCF equations may be organized into matrices according to
\begin{equation}
\label{complexeom}
i\hbar\left[
\begin{array}{ccccc}
C_{{\bf dd}}&0&\vline&C_{{\bf d}{\bm\phi}}&0\\
0&-C^*_{{\bf dd}}&\vline&0&-C^*_{{\bf d}{\bm\phi}}\\
&&\vline&&\\ 
\hline&&\vline&&\\
C^\dagger_{{\bf d}{\bm\phi}}&0&\vline&C_{{\bm\phi}{\bm\phi}}&0\\
0&-C^T_{{\bf d}{\bm\phi}}&\vline&0&-C^*_{{\bm\phi}{\bm\phi}}\\
\end{array}
\right]
\left[
\begin{array}{c}
\dot{\bf d}\\
\dot{\bf d}^*\\
\dot{\bm\phi}\\
\dot{\bm\phi}^*\\
\end{array}
\right]
=
\left[
\begin{array}{c}
\partial E/\partial{\bf d}^*\\
\partial E/\partial{\bf d}\\
\partial E/\partial{\bm\phi}^*\\
\partial E/\partial{\bm\phi}\\
\end{array}
\right].
\end{equation}
They are of the canonical form 
\begin{equation}
\omega\dot\eta=\frac{\partial H}{\partial\eta},
\end{equation}
where $\eta$ is a column vector whose entries are the dynamical variables, {\it i.e.}, $\eta=[{\bf d},{\bf d}^*,{\bm\phi},{\bm\phi}^*].$ The matrix
\begin{equation}
\label{eom3}
\omega=
\left[
\begin{array}{cc}
D&J\\
J^\dagger&M
\end{array}
\right]
=i\hbar\left[
\begin{array}{ccccc}
C_{{\bf dd}}&0&\vline&C_{{\bf d}{\bm\phi}}&0\\
0&-C^*_{{\bf dd}}&\vline&0&-C^*_{{\bf d}{\bm\phi}}\\
&&\vline&&\\ 
\hline&&\vline&&\\
C^\dagger_{{\bf d}{\bm\phi}}&0&\vline&C_{{\bm\phi}{\bm\phi}}&0\\
0&-C^T_{{\bf d}{\bm\phi}}&\vline&0&-C^*_{{\bm\phi}{\bm\phi}}\\
\end{array}
\right],
\end{equation}
which multiplies the column of velocities on the left hand side of Eq. (\ref{eom2}), is called the symplectic form \cite{Goldstein,Jose}. It is, in this case, a nonlinear complex-valued function of the dynamical variables, {\it i.e.}, $\omega\equiv\omega[{\bf d}^*,{\bm\phi}^*;{\bf d},{\bm\phi}],$ and defines the symplectic structure of the generalized phase space of the dynamical system \cite{Abraham}. It contains the unique matrix elements:
\begin{subequations}
\begin{eqnarray}
{[C_{{\bf dd}}]_{N_1N_1'}}&\equiv&{C_{d_{N_1}d_{N_1'}}}=i1_{N_1N_1'}\\
{[C_{{\bf d}{\bm\phi}}]_{N_1k}}&\equiv&{C_{d_{N_1}\phi_{k}}}=N_k\phi_{k}^*d_{N_1}\\
{[C_{{\bm\phi}{\bm\phi}}]_{kl}}&\equiv&{C_{\phi_k\phi_l}}\\
&=&\sum_{N_1=0}^Nd^*_{N_1}[N_{l}\delta_{kl}+N_{l}(N_k-\delta_{kl})\phi_{k}\phi^*_{l}]d_{N_1}\nonumber
\end{eqnarray}
\end{subequations}
For later simplification, the Fock state expansion coefficient sector of $\omega$ is blocked into the $2(N+1)\times2(N+1)$-dimensional matrix $D,$ the upper right rectangular $2(N+1)\times4K$-dimensional matrix $J$ provides the coupling between the expansion coefficients $\bf d$ and the underlying orbitals ${\bm\phi},$ and the orbital sector of $\omega$ is blocked into the $4K\times4K$-dimensional matrix $M.$ It is noted, that much of the nontrivial coupling between dynamical variables is contained in $\omega.$ The remaining coupling occurs in forces, as will now be demonstrated.

\subsection{Generalized forces}
Appearing on the right hand side of Eq. (\ref{eom1}) are the generalized forces
\begin{widetext}
\begin{equation}
\label{force1}
\begin{split}
&\frac{\partial E}{\partial d^*_{N_1}(t)}=\sqrt{(N_1+2)(N_1+1)(N_2-1)N_2}(1/2)V_{2211}(t)d_{N_1+2}(t)\\
&\ +\Bigl\{\sqrt{(N_1+1)N_2}[h_{21}(t)+V_{2111}(t)N_1+V_{2221}(t)(N_2-1)]\Bigr\}d_{N_1+1}(t)\\
&\ +\Bigl\{h_{11}(t)N_1+h_{22}(t)N_2+(1/2)[\bar V_{1212}(t)+\bar V_{2121}(t)]N_1N_2+(1/2)V_{1111}(t)(N_1^2-N_1)+(1/2)V_{2222}(t)(N_2^2-N_2)\Bigr\}d_{N_1}(t)\\
&\ +\Bigl\{\sqrt{N_1(N_2+1)}[h_{12}(t)+V_{1112}(t)(N_1-1)+V_{1222}(t)N_2]\Bigr\}d_{N_1-1}(t)\\
&\ +\sqrt{N_1(N_1-1)(N_2+1)(N_2+2)}(1/2)V_{1122}(t)d_{N_1-2}(t)
\end{split}
\end{equation}
\begin{equation}
\label{force2}
\begin{split}
\frac{\partial E}{\partial\phi_k^*({\bf x},t)}&=[h({\bf x})\gamma_{kk}(t)+{\cal V}_{kk}({\bf x},t)\Gamma_{kkkk}(t)+{\cal V}_{k'k'}({\bf x},t)\Gamma_{kk'kk'}(t)+{\cal V}_{kk'}({\bf x},t)\Gamma_{kkkk'}(t)+{\cal V}_{k'k}({\bf x},t)\Gamma_{k'kkk}(t)]\phi_k({\bf x},t)\\
&+[h({\bf x})\gamma_{kk'}(t)+{\cal V}_{kk}({\bf x},t)\Gamma_{kkkk'}(t)+{\cal V}_{k'k'}({\bf x},t)\Gamma_{kk'k'k'}(t)+{\cal V}_{kk'}({\bf x},t)\Gamma_{kkk'k'}(t)+{\cal V}_{k'k}({\bf x},t)\Gamma_{kk'k'k}(t)]\phi_{k'}({\bf x},t)
\end{split}
\end{equation}
\end{widetext}
for $N_1=0,\ldots,N$ and $k\neq k'=1,2.$ Here, the complex conjugate forces are omitted for brevity. The forces are written in terms of the Fock space matrix elements $\gamma_{kl}$ and $\Gamma_{klmn}$ of the complex-valued one- and two-body reduced density matrices $\gamma(1,1')=\langle\hat\Psi^\dagger(1)\hat\Psi(1')\rangle$ and $\Gamma(1,1';2,2')=\langle\hat\Psi^\dagger(1)\hat\Psi^\dagger(1')\hat\Psi(2)\hat\Psi(2')\rangle$ \cite{Lowdin55a,Penrose56} taken with respect to the TDMCBSCF state (\ref{ansatz}). That is
\begin{equation}
\begin{split}
&\gamma(1,1')=\langle d,\Phi[{\bm\phi}^*]|\hat\Psi^\dagger(1)\hat\Psi(1')|d,\Phi[{\bm\phi}]\rangle\\
&=\sum_{kl=1,2}\phi^*_k(1)\Bigl[\sum_{N_1N'_1=0}^Nd^*_{N_1}\langle\Phi_{N_1}[{\bm\phi}^*]|\hat b_k^\dagger\hat b_l|\Phi_{N'_1}[{\bm\phi}]\rangle d_{N'_1}\Bigr]\phi_l(1')\\
&=\sum_{kl=1,2}\phi^*_k(1)\gamma_{kl}\phi_l(1')
\end{split}
\end{equation}
\begin{widetext}
\begin{equation}
\begin{split}
&\Gamma(1,1';2,2')=\langle d,\Phi[{\bm\phi}^*]|\hat\Psi^\dagger(1)\hat\Psi^\dagger(1')\hat\Psi(2)\hat\Psi(2')|d,\Phi[{\bm\phi}]\rangle\\
&=\sum_{klmn=1,2}\phi^*_k(1)\phi^*_l(1')\Bigl[\sum_{N_1N'_1=0}^Nd^*_{N_1}\langle\Phi_{N_1}[{\bm\phi}^*]|\hat b_k^\dagger\hat b_l^\dagger\hat b_m\hat b_n|\Phi_{N'_1}[{\bm\phi}]\rangle d_{N'_1}\Bigr]\phi_m(2)\phi_n(2')\\
&=\sum_{klmn=1,2}\phi^*_k(1)\phi^*_l(1)\Gamma_{klmn}\phi_m(2)\phi_n(2')
\end{split}
\end{equation}
\end{widetext}
for $k,l,m,n=1,2.$ Note that both $\gamma$ and $\Gamma$ are nonlinear functions of the dynamical variables. The bosonic field operators $\hat\Psi({\bf x},t)$ and $\hat\Psi^\dagger({\bf x},t)$ appearing in these equations, satisfy standard boson commutation relations.

One-body matrix elements of the atom-atom interaction potential are given by ${\cal V}_{kl}({\bf x},t)=\int\phi^*_k({\bf x}',t)V({\bf x},{\bf x}')\phi_l({\bf x}',t)d^3x'.$ The symmetrized, direct plus exchange, matrix elements of $V({\bf x},{\bf x}')$ are written as $\bar V_{klmn}=V_{klmn}+V_{klnm}$ for simplicity. All of the above matrix elements, both in Fock space and in configuration space, enjoy symmetry relations among the indices, {\it e.g.}, $h_{kl}=h^*_{lk},$ ${\cal V}_{kl}={\cal V}_{lk}^*,$ $V_{klmn}=V_{lknm}=V^*_{mnkl}=V^*_{nmlk},$ and $\gamma_{kk'}=\gamma^*_{k'k}$ ($k\neq k'=1,2$). The symmetries of the Fock space matrix elements of the Hermitian two-body reduced density matrix
\begin{equation}
\label{G}
\Gamma=
\left[
\begin{array}{cccc}
\Gamma_{kkkk}&\Gamma_{kkkk'}&\Gamma_{kkk'k}&\Gamma_{kkk'k'}\\
\cdot&\Gamma_{kk'kk'}&\Gamma_{kk'k'k}&\Gamma_{kk'k'k'}\\
\cdot&\cdot&\Gamma_{k'kk'k}&\Gamma_{k'kk'k'}\\
\cdot&\cdot&\cdot&\Gamma_{k'k'k'k'}\\
\end{array}
\right]
\end{equation}
reveal that only six out of the ten matrix elements in the upper triangle of Eq. (\ref{G}) are independent; they are $\Gamma_{kkkk},$ $\Gamma_{k'k'k'k'},$ $\Gamma_{kk'kk'},$ $\Gamma_{kkk'k'},$ $\Gamma_{kkkk'},$ and $\Gamma_{kk'k'k'}.$ A similar situation exists for fermions with the labels $k$ and $k'$ replaced by holes and particles; see, {\it e.g.}, Ref. \cite{Weiner91}.

It is now evident that the TDVP approximation of the exact time-dependent many-boson Schr\"odinger equation leads to dynamical equations that are highly nonlinear both through the symplectic form $\omega$ and through the forces $\partial H/\partial\eta,$ which are themselves dynamical functions. Linearization of these multiconfigurational TDVP equations forms the basis for a bosonic multiconfigurational random phase approximation; see, {\it e.g.}, Ref. \cite{Dalgaard80} for the multiconfigurational fermionic case.

\subsection{Symplectic transformation to real-valued dynamical coordinates for numerical efficiency}
In this work, preference is given to numerical routines that involve real-valued equations and their solutions rather than their complex-valued analogs. In effort to achieve this partiality, a symplectic transformation is constructed to map Eqs. (\ref{eom3}) into manifestly real form. A symplectic (canonical) transformation is a mapping from a set of dynamical variables to a new set of dynamical coordinates that is constructed in such a way as to preserve the symplectic structure of the generalized phase space \cite{Abraham}. The transformation 
\begin{equation}
T=\frac{1}{\sqrt{2}}
\left[
\begin{array}{cc}
X&0\\
0&Y
\end{array}
\right],
\end{equation}
where both the $2(N+1)\times2(N+1)$-dimensional $X$ and the $4K\times4K$-dimensional $Y$ matrices are of the block form
\begin{equation}
\left[
\begin{array}{cc}
\textrm{diag}(1)&\textrm{diag}(1)\\
\textrm{diag}(-i)&\textrm{diag}(i)
\end{array}
\right],
\end{equation}
is constructed to map $(\xi,\xi^*)\to(\textrm{Re}\{\xi\},\textrm{Im}\{\xi\})$ and, simultaneously, preserve the structure of the dynamical equations. That is, under $T,$ the equations of motion (\ref{eom3}) become
\begin{equation}
\omega\dot\eta=\frac{\partial H}{\partial\eta}\ \to\ T\omega T^\dagger T\dot\eta=T\frac{\partial H}{\partial\eta}.
\end{equation}
The latter equation may be written more compactly as
\begin{equation}
\label{realeom}
(T\omega T^\dagger)\dot{\tilde\eta}=\tilde\omega\dot{\tilde\eta}=\frac{\partial\tilde H}{\partial\tilde\eta},
\end{equation}
where $\tilde\eta=T\eta.$ These transformed dynamical equations are manifestly real valued. The new phase space coordinates and derivatives are
\begin{equation}
\tilde\eta=[{\bf d}^R,{\bf d}^I,{\bm\phi}^R,{\bm\phi}^I]=\sqrt{2}[\textrm{Re}\{{\bf d}\},\textrm{Im}\{{\bf d}\},\textrm{Re}\{{\bm\phi}\},\textrm{Im}\{{\bm\phi}\}]
\end{equation}
and
\begin{equation}
\partial/\partial\tilde\eta=[\partial/\partial{\bf d}^R,\partial/\partial{\bf d}^I,\partial/\partial{\bm\phi}^R,\partial/\partial{\bm\phi}^I],
\end{equation}
and the new antisymmetric (or skew-symmetric) symplectic form is
\begin{equation}
\label{Ced1}
\tilde\omega=\frac{\hbar}{2}
\left[
\begin{array}{cc}
XDX^\dagger&XJY^\dagger\\
YJ^\dagger X^\dagger&YMY^\dagger
\end{array}
\right].
\end{equation}
A detailed illustration of each block of $\tilde\omega$ is now provided. The upper left $2(N+1)\times2(N+1)$-dimensional block becomes
\begin{equation}
\frac{1}{2}XDX^\dagger=
\left[
\begin{array}{cc}
0&\textrm{diag}(-1)\\
\textrm{diag}(1)&0
\end{array}
\right],
\end{equation}
while the upper right rectangular $2(N+1)\times4K$-dimensional block is
\begin{equation}
\label{rect}
\begin{split}
\frac{1}{2}[XJY^\dagger]_{N_1;lq'}&=
\left[
\begin{array}{cc}
-\textrm{Im}\Bigl(\dfrac{\partial\gamma_{ll}}{\partial d^*_{N_1}}\phi^*_{lq'}\Bigr)&-\textrm{Re}\Bigl(\dfrac{\partial\gamma_{ll}}{\partial d^*_{N_1}}\phi^*_{lq'}\Bigr)\\
\textrm{Re}\Bigl(\dfrac{\partial\gamma_{ll}}{\partial d^*_{N_1}}\phi^*_{lq'}\Bigr)&-\textrm{Im}\Bigl(\dfrac{\partial\gamma_{ll}}{\partial d^*_{N_1}}\phi^*_{lq'}\Bigr)
\end{array}
\right]\\
&=\left[
\begin{array}{cc}
-N_l\textrm{Im}(d_{N_1}\phi^*_{lq'})&-N_l\textrm{Re}(d_{N_1}\phi^*_{lq'})\\
N_l\textrm{Re}(d_{N_1}\phi^*_{lq'})&-N_l\textrm{Im}(d_{N_1}\phi^*_{lq'})
\end{array}
\right]
\end{split}
\end{equation}
and the lower right $4K\times4K$-dimensional block is given by
\begin{equation}
\label{AB}
\frac{1}{2}YMY^\dagger=
\left[
\begin{array}{cc}
A&B\\
-B^T&A
\end{array}
\right],
\end{equation}
in terms of the $2K\times2K$-dimensional sub-blocks
\begin{widetext}
\begin{equation}
\label{Ced3}
\begin{split}
A_{kq;lq'}&=\left[
\begin{array}{cc}
-\Gamma_{1111}\textrm{Im}(\phi_{1q}\phi^*_{1q'})&-\Gamma_{1212}\textrm{Im}(\phi_{1q}\phi^*_{2q'})\\
-\Gamma_{2121}\textrm{Im}(\phi_{2q}\phi^*_{1q'})&-\Gamma_{2222}\textrm{Im}(\phi_{2q}\phi^*_{2q'})
\end{array}
\right]\\
&=
\left[
\begin{array}{cc}
-\sum_{N_1=0}^N |d_{N_1}|^2N_1(N_1-1)\textrm{Im}(\phi_{1q}\phi^*_{1q'})&-\sum_{N_1=0}^N |d_{N_1}|^2N_1N_2\textrm{Im}(\phi_{1q}\phi^*_{2q'})\\
-\sum_{N_1=0}^N |d_{N_1}|^2N_2N_1\textrm{Im}(\phi_{2q}\phi^*_{1q'})&-\sum_{N_1=0}^N |d_{N_1}|^2N_2(N_2-1)\textrm{Im}(\phi_{2q}\phi^*_{2q'})
\end{array}
\right]\\
B_{kq;lq'}&=
\left[
\begin{array}{cc}
-[\gamma_{11}\delta_{qq'}+\Gamma_{1111}\textrm{Re}(\phi_{1q}\phi^*_{1q'})]&-\Gamma_{1212}\textrm{Re}(\phi_{1q}\phi^*_{2q'})\\
-\Gamma_{2121}\textrm{Re}(\phi_{2q}\phi^*_{1q'})&-[\gamma_{22}\delta_{qq'}+\Gamma_{2222}\textrm{Re}(\phi_{2q}\phi^*_{2q'})]
\end{array}
\right]\\
&=
\left[
\begin{array}{cc}
-\sum_{N_1=0}^N |d_{N_1}|^2[N_1\delta_{qq'}+N_1(N_1-1)\textrm{Re}(\phi_{1q}\phi^*_{1q'})]&-\sum_{N_1=0}^N|d_{N_1}|^2N_1N_2\textrm{Re}(\phi_{1q}\phi^*_{2q'})\\
-\sum_{N_1=0}^N|d_{N_1}|^2N_2N_1\textrm{Re}(\phi_{2q}\phi^*_{1q'})&-\sum_{N_1=0}^N |d_{N_1}|^2[N_2\delta_{qq'}+N_2(N_2-1)\textrm{Re}(\phi_{2q}\phi^*_{2q'})]
\end{array}
\right].
\end{split}
\end{equation}
\end{widetext}
In both Eqs. (\ref{rect}) and (\ref{AB}), the contraction is on $l=1,2$ and $q'=1,\ldots,K.$ The lower left rectangular $4K\times2(N+1)$-dimensional block of $\tilde\omega$ is the negative transpose of Eq. (\ref{rect}).

The transformed real-valued forces are simply related to the real and imaginary parts of Eqs. (\ref{force1}) and (\ref{force2}). That is
\begin{equation}
\begin{split}
\frac{\partial E}{\partial {\bf d}^R}&=\frac{1}{\sqrt{2}}\Bigl[\frac{\partial E}{\partial {\bf d}^*}+\frac{\partial E}{\partial {\bf d}}\Bigr]\\
\frac{\partial E}{\partial {\bf d}^I}&=\frac{1}{\sqrt{2}}\Bigl[-i\frac{\partial E}{\partial {\bf d}^*}+i\frac{\partial E}{\partial {\bf d}}\Bigr]\\
\frac{\partial E}{\partial{\bm\phi}^R}&=\frac{1}{\sqrt{2}}\Bigl[\frac{\partial E}{\partial{\bm\phi}^*}+\frac{\partial E}{\partial{\bm\phi}}\Bigr]\\
\frac{\partial E}{\partial{\bm\phi}^I}&=\frac{1}{\sqrt{2}}\Bigl[-i\frac{\partial E}{\partial{\bm\phi}^*}+i\frac{\partial E}{\partial{\bm\phi}}\Bigr].
\end{split}
\end{equation}
As before, once the initial values for $\tilde\eta$ are specified, then the real-valued TDMCBSCF equations (\ref{realeom}) form a well-defined and numerically efficient initial value problem, that is equivalent to the complex-valued equations (\ref{complexeom}), and contain the complete coupling between the dynamical coordinates parametrizing the TDVP state vector (\ref{ansatz}). The real-valued phase space, in which the solution of Eq. (\ref{realeom}) evolves, is endowed with the Poisson bracket 
\begin{equation}
\{F,G\}=\frac{\partial F}{\partial\tilde\eta^T}\tilde\omega^{-1}\frac{\partial G}{\partial\tilde\eta}
\end{equation}
so that, together with the symplectic structure generated by $\{\cdot,\cdot\},$ the TDMCBSCF representation of the exact many-body Schr\"odinger equation takes on the classical Hamiltonian form 
\begin{equation}
\dot{\tilde\eta}=\{\tilde\eta,\tilde H\}.
\end{equation}
These dynamical equations have been completely implemented in an extensive computer program from which the first numerical results will now be presented.

\section{Numerical Implementation}
Together with an initial value for the vector $\eta,$ the above set of first order Hamiltonian (TDVP) dynamical equations is well defined. A simplistic quadrature scheme such as 
\begin{equation}
\label{Euler}
\eta(t+\varepsilon)=\eta(t)+\varepsilon\dot\eta(t)=\eta(t)+\varepsilon\omega^{-1}(t)\frac{\partial H}{\partial\eta(t)}
\end{equation}
may be written to integrate the state vector forward in time from $t$ to $t+\varepsilon$ \cite{fn11}. This does require that the symplectic form $\omega$ be inverted at each time step; however, it has been found that, in practice, $\omega$ is readily inverted with standard $AX=B$ inversion routines \cite{lapack}. These routines are easily incorporated into numerical integrators \cite{Press1} that are more sophisticated than Eq. (\ref{Euler}). This matter will be discussed in greater detail in the following.

\subsection{Preliminaries}
The TDMCBSCF evolution equations (\ref{realeom}), generated from the TDVP, have, so far, been represented in an arbitrary basis of rank $K$ of functions $g_q({\bf x}).$ In practice, these equations are solved using a fast Fourier transform based pseudospectral grid method \cite{fftw,Gottlieb} in quasi-one dimension \cite{Carr00,Olshanii1998a}. The Fourier grid basis is constructed so that the basis functions $g_q({\bf x})$ enforce the appropriate boundary conditions on the orbitals at the grid boundaries. For the box boundary conditions employed here, these basis functions are the Fourier sine functions. It is found that the solutions of Eq. (\ref{realeom}) are adequately converged with $K=2^9$ fixed grid points in quasi-one dimension. For the short time dynamics presented in the following, this value of $K$ is sufficiently large enough to avoid the case where the BEC density is appreciably different from zero at the box edges. Propagation to longer times, on the order of 50 ms or larger, may require the use of more grid points. The extension to three dimensions poses a challenge that may be overcome, without difficulty, with increased computing time.

The equations of motion (\ref{realeom}) are integrated forward in time in real space, where a Fourier transform to $k$-space is made at each time step to diagonalize the spatial derivatives appearing in the generalized forces (\ref{force1}) and (\ref{force2}). These terms are then transformed back to direct space before time propagation. Inversion of the symplectic form $\tilde\omega$ is achieved, in direct space, with standard linear algebra routines \cite{lapack} and is incorporated into a variable step size fourth order Runge-Kutta method adopted from Ref. \cite{Press1}. Difficulties arising from the inversion of $\tilde\omega,$ which is on the order of several thousand by several thousand, have yet to be encountered; however, analytical and numerical methods do exist to correct this situation should it arise in the future \cite{Kan81,Deumens94}. These methods rely on Darboux's theorem (see, {\it e.g.}, Ref. \cite{Abraham}) to find a local coordinate chart in a curved phase space where $\tilde\omega$ takes on the constant form
\begin{equation}
\tilde\omega\longrightarrow
\left[
\begin{array}{cc}
0&I\\
-I&0
\end{array}
\right].
\end{equation}
In a flat phase space, it is always possible to find a global system of coordinates where this result holds. Since the TDMCBSCF phase space is curved, transformation of $\tilde\omega$ to constant form is only possible locally.

Values of the physical parameters such as the atomic mass $m,$ scattering length $a,$ and oscillator length $\beta=\sqrt{\hbar/m\omega},$ which enter the Hamiltonian (\ref{ham}), are taken from Ref. \cite{Masiello05}, as is the functional form of the external trapping potential $V_{\textrm{ext}}.$ These values are scaled, in the quasi-one-dimensional approximation \cite{Carr00,Olshanii1998a}, so that the product $aN,$ with $N=100,$ is consistent with the $^{23}$Na double-well interference experiments \cite{Shin2004a,Shin2004b,Saba05} performed at MIT. It is these experiments, and those in Heidelberg \cite{Albiez04}, that have inspired the following numerical examples. Note that the TDMCBSCF theory is not limited to the quasi-one-dimensional approximation, the contact interaction approximation, nor to the particular choice of $V_{\textrm{ext}}$ that is used in the following.

\subsection{Numerical Results}
Imaging of typical laboratory BECs is commonly performed by releasing the condensate's trapping potential, and, subsequently, allowing the degenerate gas to ballistically expand to a point where it is large enough to be photographed. When atoms are, {\it e.g.}, condensed into a double-well trap, release of the trap enables the initially left- and right-localized atomic clouds, which may or may not be coherent, to expand and eventually overlap due to their mutually repulsive interactions \cite{Andrews1997b}. The details of the overlap dynamics depend strongly upon the initial phase coherence of the BEC and, additionally, upon the degree of atomic correlation between atoms in different Fock states.

Using the TDMCBSCF theory, we present the ballistic expansion dynamics of two initially stationary states of the double-well condensate of $N$ identical atoms at zero temperature, where, in each case, the left- and right-localized condensate moieties are:
\begin{itemize}
\item[(1.)] 
initially stationary and, consequently, phase coherent across both wells
\item[(2.)] 
initially phase offset in each orbital by $(\pi/4)/N_1$ in the left hand well
\end{itemize}
The two stationary states of interest are the fragmented ground state and a high-lying macroscopic quantum superposition state of the BEC. The latter is a highly entangled number state and is also called a Schr\"odinger cat. Both are obtained, initially, from full time-independent MCBSCF calculations; see Refs. \cite{Masiello05,Masiello06}. Following the release of the trap at $t=0,$ neither of these states are stationary. It is their many-body density
\begin{equation}
\begin{split}
\rho&({\bf x},t)\equiv\gamma({\bf x},t;{\bf x}'={\bf x},t'=t)\\
&=\sum_{N_1=0}^N|d_{N_1}(t)|^2[N_1|\phi_1({\bf x},t)|^2+N_2|\phi_2({\bf x},t)|^2]\\
&\ +\sum_{N_1=1}^N\sqrt{N_1(N_2+1)}d^{*}_{N_1}(t)\phi_1^*({\bf x},t)\phi_2({\bf x},t)d_{N_1-1}(t)\\
&\ +\sum_{N_1=0}^{N-1}\sqrt{(N_1+1)N_2}d^{*}_{N_1}(t)\phi_2^*({\bf x},t)\phi_1({\bf x},t)d_{N_1+1}(t),
\end{split}
\end{equation}
orbitals $\phi_k({\bf x},t),$ and expansion coefficients $d_{N_1}(t)$ that are followed as functions of space and time.

\begin{figure*}
\psfrag{x}[][]{{\large $x$ $(\beta)$}}
\psfrag{rho}[][]{density $(\beta^{-1})$}
\psfrag{-30}[][]{{-30}}
\psfrag{-20}[][]{{-20}}
\psfrag{-10}[][]{{-10}}
\psfrag{ 0}[][]{{0}}
\psfrag{ 30}[][]{{30}}
\psfrag{ 20}[][]{{20}}
\psfrag{ 10}[][]{{10}}
\psfrag{ 4}[][]{{4}}
\psfrag{ 3}[][]{{3}}
\psfrag{ 2}[][]{{2}}
\psfrag{ 1}[][]{{1}}
\psfrag{p0}[][]{{\hspace{1cm}stationary state}}
\psfrag{p1}[][]{{\hspace{1cm}$\dfrac{\pi/4}{N/2}$ offset}}
\psfrag{t1}[][]{{$t=1$ ms}}
\psfrag{t2}[][]{{$t=2$ ms}}
\psfrag{t3}[][]{{$t=2$ ms}}
\rotatebox{0}{\resizebox{!}{10cm}{\includegraphics{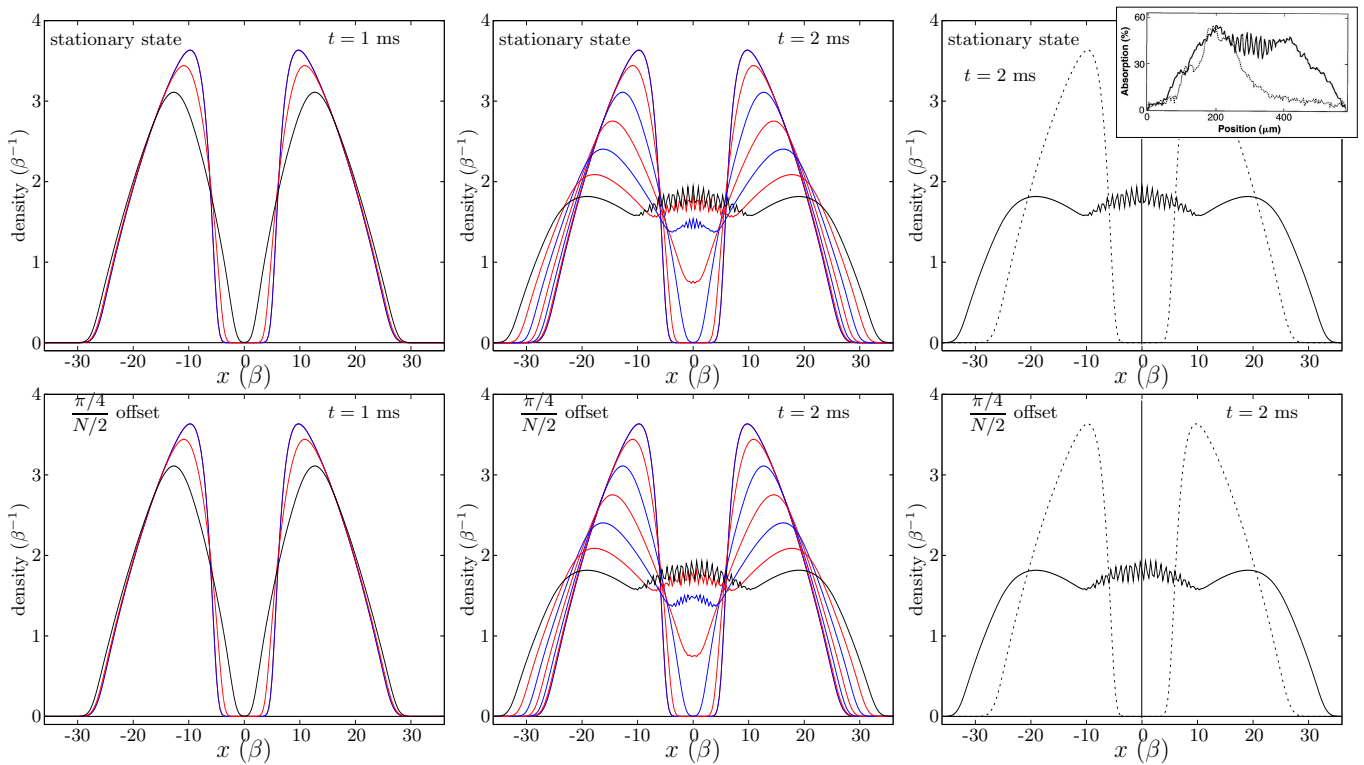}}}
\caption{\label{frag1}(Color online) Ballistic expansion dynamics of the initially fragmented ground state density of the double-well condensate computed within the TDMCBSCF theory. The upper three panels display the dynamics of an initially stationary state of the BEC, while the lower three panels display that of an initially $(\pi/4)/(N/2)$ orbital phase offset between the left and right wells. Time increases from left to right, with snapshots taken every 0.25 ms, so that a history of the dynamics is captured at 1 ms and 2 ms after release of the trap at $t=0.$ It might appear that only three snapshots exist in frame one, but the color version will show that the density at $t=0.25$ ms overlaps the initial density. Colors alternate to aid in visualization. Here, the short time dynamics reveals clearly visible interference between the left and right pieces of an initially fragmented condensate. The location of the interference fringes are not random since, in TDMCBSCF theory, the initial fragments have a well-defined phase relationship. The insert in panel three is taken, with permission pending, from the early BEC interference experiments performed at MIT where two independent condensates were allowed to ballistically expand and overlap; see Ref. \cite{Andrews1997b}. Note that the experimentally observed interference fringes do not have 100$\%$ visibility in agreement with our TDMCBSCF theory. It will be demonstrated that Gross-Pitaevskii theory vastly overemphasizes the visibility of these fringes.}
\end{figure*}

\subsubsection{Ballistic expansion dynamics of an initially fragmented state}
\begin{figure*}
\psfrag{x}[][]{{\large $x$ $(\beta)$}}
\psfrag{HF orbitals1}[][]{$\phi_1({\bf x},t)\ $ $(\beta^{-1/2})$}
\psfrag{HF orbitals2}[][]{$\phi_2({\bf x},t)\ $ $(\beta^{-1/2})$}
\psfrag{CN1}[][]{{$d_{N_1}(t)$}}
\psfrag{N1}[][]{{$N_1/N$}}
\psfrag{-10}[][]{{-10}}
\psfrag{-20}[][]{{-20}}
\psfrag{-30}[][]{{-30}}
\psfrag{ 0c}[][]{{0}}
\psfrag{ 0c0}[][]{{}}
\psfrag{ 1c}[][]{{1}}
\psfrag{ 0.2c}[][]{{0.2}}
\psfrag{ 0.4c}[][]{{0.4}}
\psfrag{ 0.6c}[][]{{0.5}}
\psfrag{ 0.8c}[][]{{0.8}}
\psfrag{ 0.1c}[][]{{}}
\psfrag{ 0.3c}[][]{{}}
\psfrag{ 0.5c}[][]{{}}
\psfrag{ 0.7c}[][]{{}}
\psfrag{ 0.9c}[][]{{}}
\psfrag{ 30}[][]{{30}}
\psfrag{ 20}[][]{{20}}
\psfrag{ 10}[][]{{10}}
\psfrag{ 22c}[][]{{\rotatebox{90}{3 ms}}}
\psfrag{ 20c}[][]{{}}
\psfrag{ 18c}[][]{{}}
\psfrag{ 16c}[][]{{}}
\psfrag{ 17}[][]{{}}
\psfrag{ 16}[][]{{\rotatebox{90}{2 ms}}}
\psfrag{ 15}[][]{{}}
\psfrag{ 14}[][]{{}}
\psfrag{ 14c}[][]{{\rotatebox{90}{2 ms}}}
\psfrag{ 13}[][]{{}}
\psfrag{ 12}[][]{{}}
\psfrag{ 12c}[][]{{}}
\psfrag{ 11}[][]{{}}
\psfrag{ 10}[][]{{10}}
\psfrag{ 10a}[][]{{}}
\psfrag{ 10c}[][]{{}}
\psfrag{ 9}[][]{{}}
\psfrag{ 8}[][]{{\rotatebox{90}{1 ms} }}
\psfrag{ 8c}[][]{{}}
\psfrag{ 7}[][]{{}}
\psfrag{ 6}[][]{{}}
\psfrag{ 6c}[][]{{\rotatebox{90}{1 ms}}}
\psfrag{ 5}[][]{{}}
\psfrag{ 4c}[][]{{}}
\psfrag{ 4}[][]{{}}
\psfrag{ 3}[][]{{}}
\psfrag{ 2c}[][]{{}}
\psfrag{ 2}[][]{{}}
\psfrag{ 1}[][]{{}}
\psfrag{ 0}[][]{{0}}
\psfrag{p0}[][]{{\hspace{1cm}stationary state}}
\psfrag{p1}[][]{{\hspace{1cm}$\dfrac{\pi/4}{N/2}$ offset}}
\rotatebox{0}{\resizebox{!}{10cm}{\includegraphics{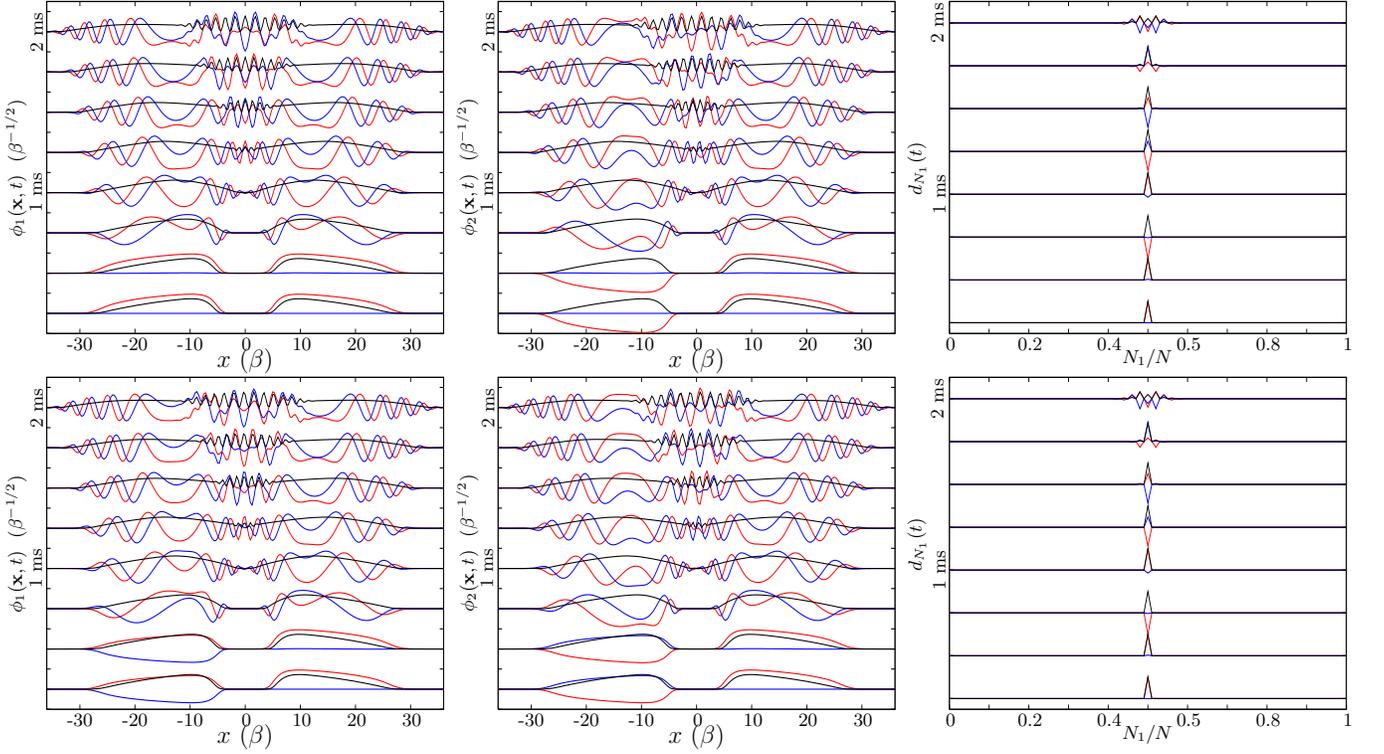}}}
\caption{\label{frag_orb1}(Color online) Orbitals and expansion coefficients corresponding to the ballistic expansion of an initially fragmented ground state of the double-well condensate computed within the TDMCBSCF theory. The upper three panels display the dynamics of an initially stationary state of the BEC, while the lower three panels display that of an initially $(\pi/4)/(N/2)$ orbital phase offset between the left and right wells. Time increases from bottom to top in each panel, with snapshots taken every 0.25 ms, so that a history of the dynamics is captured at 1 ms and 2 ms after release of the trap at $t=0.$ Colors alternate to aid in visualization with red, blue, and black representing the real, imaginary and square modulus of the fragmented state orbitals and expansion coefficients. The units of the orbitals and expansion coefficients are arbitrary.}
\end{figure*}

The fragmented TDMCBSCF state is initially single-configurational, {\it i.e.}, it involves only the single configuration 
\begin{equation}
|{\bf d};{\bm\phi}\rangle=|\Phi_{N/2}[{\bm\phi}]\rangle\equiv|N/2,N/2\rangle
\end{equation}
with $N/2$ atoms in each well for $N=100.$ No correlations exist at $t=0$ between the atoms, although a specific phase relation must be chosen between left and right fragments; correlations do arise as the dynamics unfolds. Fig. \ref{frag1} displays the short time expansion dynamics of the fragmented ground state density. The dynamics of an initially stationary state is presented in the upper three panels, while that of a condensate having an initially $(\pi/4)/N_1$ orbital phase offset between the left and right wells is displayed in the lower three panels, where, in this case $N_1=N/2.$ Time increases from left to right with each snapshot separated by 0.25 ms, so that the fragmented ground state dynamics is captured at 1 ms (panel one) and 2 ms (panel two) after the release of the trap at $t=0.$ For clarity, the third panel redisplays the density snapshot at 2 ms in the foreground and the initial state density in the background. In each panel, the density at the final integration step is plotted in black. It can be seen that an initially fragmented BEC state does produce clearly visible interference fringes upon overlap of the left and right fragments. In an experiment with random relative phase between two independently prepared condensates, the location of the interference fringes would be random \cite{Naraschewski1996a,Leggett2001a,Baym06a}; however, in this theoretical model, there is always a well-defined phase relationship between the two moieties, and, consequently, the interference fringes are not randomly located. Note that the specific but natural choice of zero phase offset between left and right fragments has been assumed in the upper three panels of Figs. \ref{frag1} and \ref{frag_orb1}. The insert in panel three of Fig. \ref{frag1} is taken from the early BEC interference experiments performed at MIT where two initially independently prepared condensates were allowed to ballistically expand, overlap, and interfere; see Ref. \cite{Andrews1997b}. Note that the visibility of the experimentally observed interference fringes is not 100$\%,$ which is in agreement with our fully variational TDMCBSCF theory; a similar result is found from a first order perturbative treatment in Ref. \cite{Ced06c}. In the following, it will be shown that Gross-Pitaevskii theory, which is a single-configurational approach, greatly overemphasizes the visibility of these interference finges.

The underlying orbitals $\phi_k({\bf x},t)$ and expansions coefficients $d_{N_1}(t),$ corresponding to the initially fragmented density in Fig. \ref{frag1}, were obtained, from first principles, with our time-independent MCBSCF theory \cite{Masiello05}. In general, the construction of an arbitrary self-consistent initial state can be difficult. Here, we rely on the fact that, within our restricted model space, the state of a left and right fragmented BEC can be re-expanded as a complicated (but equivalent) multiconfigurational superposition of symmetric and antisymmetric states with varying numbers of atoms in each. Since the TDMCBSCF state is an arbitrary vector in the space spanned by either the left/right or symmetric/antisymmetric representation, it may be equivalently represented in either basis. This fact is used to ensure that the underlying symmetric and antisymmetric orbitals $\phi_k({\bf x},t)$ used here actually correspond to an initially fragmented state.

Fig. \ref{frag_orb1} presents a series of snapshots of the orbitals $\phi_k({\bf x},t)$ and expansion coefficients $d_{N_1}(t)$ of the initially fragmented TDMCBSCF state $|{\bf d};{\bm\phi}\rangle=|N/2,N/2\rangle.$ Each snapshot is separated by 0.25 ms. The expansion coefficients, which represent the distribution of atoms between the left and right wells, initially correspond to a sharp $|N/2,N/2\rangle$ single-configurational state. Following release of the external trapping potential, it can be seen that $d_{N_1}(t)$ broadens in time and begins to approach a binomial distribution. It is well known that, in two-state multiconfigurational models, the expansion coefficients corresponding to the coherent symmetric ground state take the form of a binomial distribution peaked around the $|N/2,N/2\rangle$ configuration; see, {\it e.g.}, Ref. \cite{Spekkens1999a} and Ref. \cite{Masiello05}. In light of this fact, we note that the state of the BEC is moving from a single-configurational description that is initially left and right localized in each well to a multiconfigurational delocalized description. In an appropriate basis of symmetric and antisymmetric states, the latter delocalized multiconfigurational state reduces to the single-configurational Gross-Pitaevskii (GP) solution. The initial fragmented state is well described by Hartree-Fock (HF) theory with left and right localized orbitals. Similarly, the coherent delocalized ground state is well described by GP theory with a single symmetric orbital. However, it is impossible, within either of these single-configurational approaches, to transition from the first to the second description.  For example, by allowing a left localized and a right localized orbital to expand and eventually overlap in time-dependent HF theory, no interference occurs at any time since HF theory is a noninteracting single-particle theory. Furthermore, the final state of the system, following overlap, does not settle down into a symmetric coherent state solution of the GP equation. It is especially important to note that a full time-dependent multiconfigurational theory, such as the TDMCBSCF approach, is necessary to correctly describe the interaction physics and properly transition between the two single-configurational pictures \cite{fn14}.

\begin{figure*}
\psfrag{x}[][]{{\large $x$ $(\beta)$}}
\psfrag{rho}[][]{density $(\beta^{-1})$}
\psfrag{-40}[][]{{-40}}
\psfrag{-30}[][]{{-30}}
\psfrag{-20}[][]{{-20}}
\psfrag{-10}[][]{{-10}}
\psfrag{ 0}[][]{{0}}
\psfrag{ 40}[][]{{40}}
\psfrag{ 30}[][]{{30}}
\psfrag{ 20}[][]{{20}}
\psfrag{ 10}[][]{{10}}
\psfrag{ 3}[][]{{3}}
\psfrag{ 2}[][]{{2}}
\psfrag{ 1}[][]{{1}}
\psfrag{p0}[][]{{\hspace{1cm}stationary state}}
\psfrag{p1}[][]{{\hspace{1cm}$\dfrac{\pi/4}{N_1}$ offset}}
\psfrag{t1}[][]{{$t=1$ ms}}
\psfrag{t2}[][]{{$t=2$ ms}}
\psfrag{t3}[][]{{$t=3$ ms}}
\rotatebox{0}{\resizebox{!}{10cm}{\includegraphics{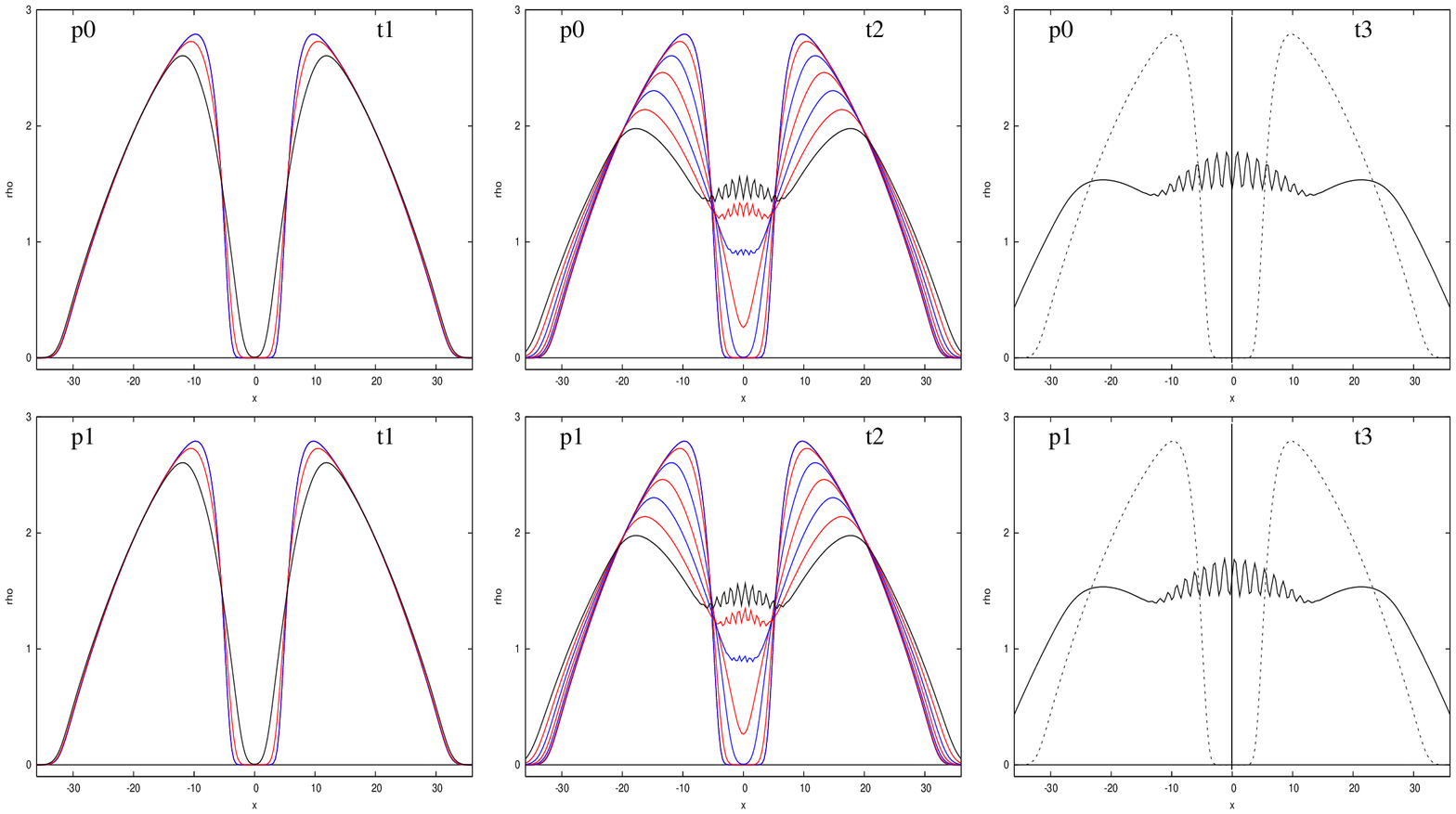}}}
\caption{\label{cat1}(Color online) Ballistic expansion dynamics of a macroscopic quantum superposition state [the entangled number state given in Eq. (\ref{cat})] of the double-well condensate computed within the TDMCBSCF theory. The upper three panels display the dynamics of an initially stationary state, while the lower three panels display that of an initially $(\pi/4)/N_1$ orbital phase offset between the left and right wells, where, in this case $N_1=90.$ Time increases from left to right, with snapshots taken every 0.25 ms so that a history of the dynamics is captured at 1 ms, 2 ms, and 3 ms after release of the trap at $t=0.$ It might appear that only three snapshots exist in frame one, but the color version will show that the density at $t=0.25$ ms overlaps the initial density. Colors alternate to aid in visualization. It is strikingly apparent that interference occurs in the short time dynamics of the superposition state density, and that the expected sensitivity to the small $(\pi/4)/N_1$ phase offset is easily observed.}
\end{figure*}

\subsubsection{Ballistic expansion dynamics of an initially number entangled state}
In order to explore the effects of initial atomic correlation on the dynamics of the double-well condensate, results are presented for the ballistic expansion of a macroscopic quantum superposition state, which is double-configurational. It represents the simplest possible example where correlations are present at $t=0.$ To this end, we choose a (superposed) number entangled state vector of the form  
\begin{equation}
\begin{split}
\label{cat}
|{\bf d};{\bm\phi}\rangle&=d_{N_1}|\Phi_{N_1}[{\bm\phi}]\rangle+d_{N_2}|\Phi_{N_2}[{\bm\phi}]\rangle\\
&=(1/\sqrt{2})|N_1,N_2\rangle+(1/\sqrt{2})|N_2,N_1\rangle
\end{split}
\end{equation}
with $N_1=90$ and $N_2=10$ for $N=100$ atoms. Fig. \ref{cat1} displays the short time expansion dynamics of the superposition in Eq. (\ref{cat}) after the trapping potential is released at $t=0.$ The dynamics of a superposition that is initially stationary is presented in the upper three panels, while that of a superposition with underlying orbitals that are initially phase offset by $(\pi/4)/(N_1-N_2)$ between the left and right wells is presented in the lower three panels, where, in this case $N_1=90\approx N\gg N_2$ and, consequently, we take $(\pi/4)/(N_1-N_2)\approx(\pi/4)/N_1\approx(\pi/4)/N.$ Therefore, the orbital phase offset of $(\pi/4)/N$ leads to a macroscopic phase offset of $\pi/4$ in Eq. (\ref{cat}). Snapshots are taken, from left to right, every 0.25 ms so that the dynamics is captured at 1 ms, 2 ms, and 3 ms following the trap's release. For clarity, the history ending at 3 ms shows only the latest update to the density, where the initial state is displayed in the background. In each panel, the density at the final integration step is plotted in black. Pronounced matter-wave interference is observed at short times (and continues to grow in at longer times), as the dynamics of the superposition state (\ref{cat}) is essentially that of the two interacting pieces of a Schr\"odinger cat, with all $N$ atoms simultaneously interfering \cite{dunningham2005}. The behavior of this superposition state is strikingly different from that of the previous fragmented state. Sensitivity to the small $(\pi/4)/N_1$ orbital phase offset is easily seen.

\begin{figure*}
\psfrag{x}[][]{{\large $x$ $(\beta)$}}
\psfrag{HF orbitals1}[][]{$\phi_1({\bf x},t)\ $ $(\beta^{-1/2})$}
\psfrag{HF orbitals2}[][]{$\phi_2({\bf x},t)\ $ $(\beta^{-1/2})$}
\psfrag{CN1}[][]{{$d_{N_1}(t)$}}
\psfrag{N1}[][]{{$N_1/N$}}
\psfrag{-10}[][]{{-10}}
\psfrag{-20}[][]{{-20}}
\psfrag{-30}[][]{{-30}}
\psfrag{ 0c}[][]{{0}}
\psfrag{ 0c0}[][]{{}}
\psfrag{ 1c}[][]{{1}}
\psfrag{ 0.2c}[][]{{0.2}}
\psfrag{ 0.4c}[][]{{0.4}}
\psfrag{ 0.6c}[][]{{0.5}}
\psfrag{ 0.8c}[][]{{0.8}}
\psfrag{ 0.1c}[][]{{}}
\psfrag{ 0.3c}[][]{{}}
\psfrag{ 0.5c}[][]{{}}
\psfrag{ 0.7c}[][]{{}}
\psfrag{ 0.9c}[][]{{}}
\psfrag{ 30}[][]{{30}}
\psfrag{ 20}[][]{{20}}
\psfrag{ 10}[][]{{10}}
\psfrag{ 22c}[][]{{\rotatebox{90}{3 ms}}}
\psfrag{ 20c}[][]{{}}
\psfrag{ 18c}[][]{{}}
\psfrag{ 16c}[][]{{}}
\psfrag{ 17}[][]{{}}
\psfrag{ 16}[][]{{\rotatebox{90}{2 ms}}}
\psfrag{ 15}[][]{{}}
\psfrag{ 14}[][]{{}}
\psfrag{ 14c}[][]{{\rotatebox{90}{2 ms}}}
\psfrag{ 13}[][]{{}}
\psfrag{ 12}[][]{{}}
\psfrag{ 12c}[][]{{}}
\psfrag{ 11}[][]{{}}
\psfrag{ 10}[][]{{10}}
\psfrag{ 10a}[][]{{}}
\psfrag{ 10c}[][]{{}}
\psfrag{ 9}[][]{{}}
\psfrag{ 8}[][]{{\rotatebox{90}{1 ms} }}
\psfrag{ 8c}[][]{{}}
\psfrag{ 7}[][]{{}}
\psfrag{ 6}[][]{{}}
\psfrag{ 6c}[][]{{\rotatebox{90}{1 ms}}}
\psfrag{ 5}[][]{{}}
\psfrag{ 4c}[][]{{}}
\psfrag{ 4}[][]{{}}
\psfrag{ 3}[][]{{}}
\psfrag{ 2c}[][]{{}}
\psfrag{ 2}[][]{{}}
\psfrag{ 1}[][]{{}}
\psfrag{ 0}[][]{{0}}
\psfrag{p0}[][]{{\hspace{1cm}stationary state}}
\psfrag{p1}[][]{{\hspace{1cm}$\dfrac{\pi/4}{N/2}$ offset}}
\rotatebox{0}{\resizebox{!}{10cm}{\includegraphics{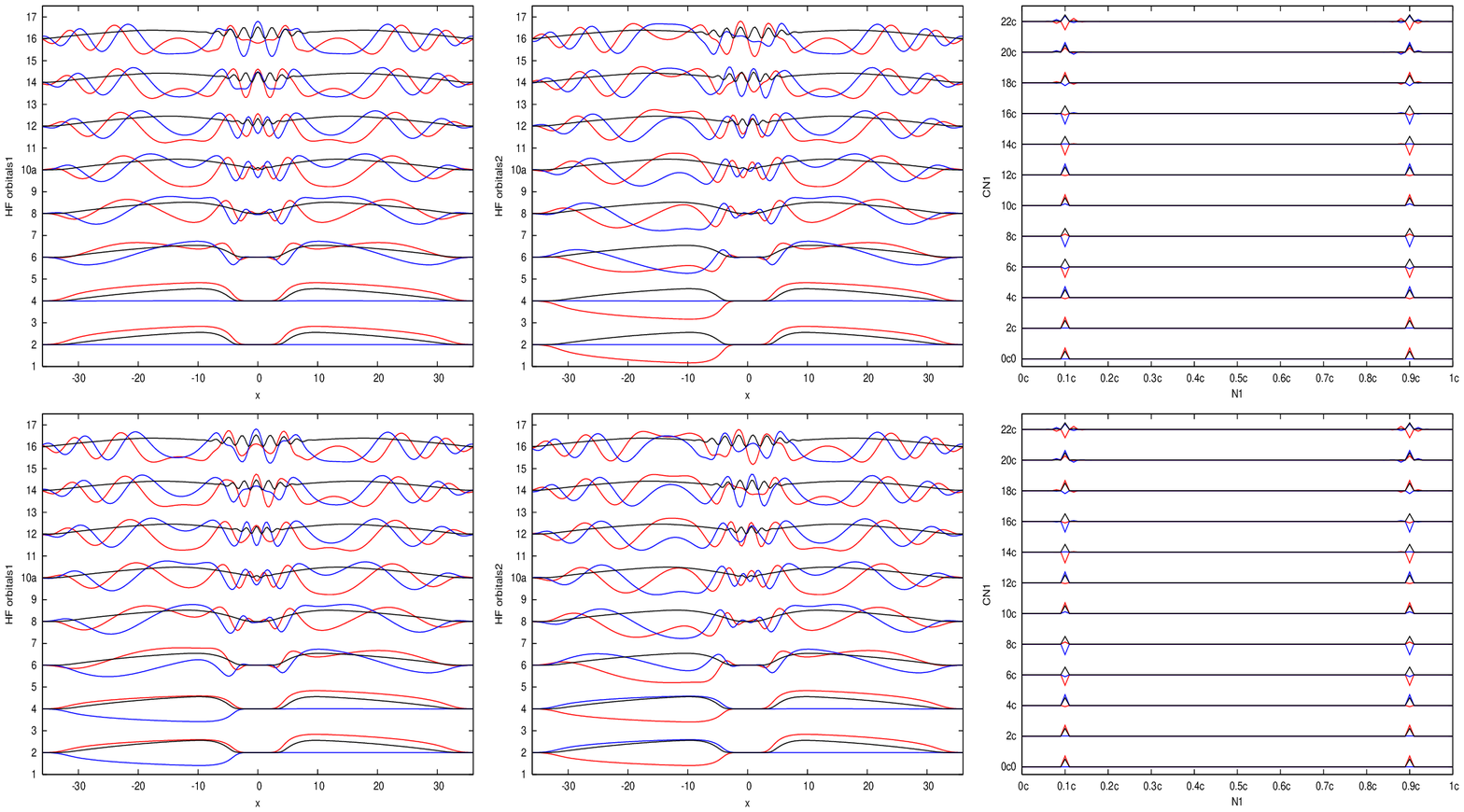}}}
\caption{\label{cat_orb1}(Color online) Orbitals and expansion coefficients corresponding to the ballistic expansion of a macroscopic quantum superposition state [the entangled number state given in Eq. (\ref{cat})] of the double-well condensate computed within the TDMCBSCF theory. The upper three panels display the dynamics of an initially stationary state, while the lower three panels display that of an initially $(\pi/4)/N_1$ orbital phase offset between the left and right wells, where, in this case $N_1=90.$ Time increases from bottom to top, with snapshots taken every 0.25 ms so that a history of the dynamics is captured at 1 ms, 2 ms, and 3 ms (for the expansion coefficients only) after release of the trap at $t=0.$ Colors alternate to aid in visualization with red, blue, and black representing the real, imaginary and square modulus of the fragmented state orbitals and expansion coefficients. The units of the orbitals and expansion coefficients are arbitrary.}
\end{figure*}

Fig. \ref{cat_orb1} presents the orbitals $\phi_k({\bf x},t)$ and expansion coefficients $d_{N_1}(t)$ that are appropriate for the number entangled state (\ref{cat}). These orbitals and expansion coefficients were obtained, from first principles, with our time-independent symmetry-breaking MCBSCF theory \cite{Masiello06}. As discussed previously, the determination of correct and mutually self-consistent initial orbitals and expansion coefficients is, by no means, an easy task. Note that, in distinction to the fragmented state orbitals corresponding to $N/2$ atoms in each well, the entangled state orbitals reflect the fact almost all $N$ atoms are in both wells simultaneously. Consequently, the entangled state orbitals are much broader than the fragmented state orbitals, due to their mutually repulsive interactions. This broadening manifests itself in the interference patterns of the number entangled state.

\begin{figure*}
\psfrag{x}[][]{{\large $x$ $(\beta)$}}
\psfrag{rho}[][]{density $(\beta^{-1})$}
\psfrag{-40}[][]{{-40}}
\psfrag{-30}[][]{{-30}}
\psfrag{-20}[][]{{-20}}
\psfrag{-10}[][]{{-10}}
\psfrag{ 0}[][]{{0}}
\psfrag{ 40}[][]{{40}}
\psfrag{ 30}[][]{{30}}
\psfrag{ 20}[][]{{20}}
\psfrag{ 10}[][]{{10}}
\psfrag{ 3}[][]{{3}}
\psfrag{ 2}[][]{{2}}
\psfrag{ 1}[][]{{1}}
\psfrag{p0}[][]{{\hspace{1cm}stationary state}}
\psfrag{p1}[][]{{\hspace{1cm}$\pi/4$ offset}}
\psfrag{t1}[][]{{$t=1$ ms}}
\psfrag{t2}[][]{{$t=2$ ms}}
\psfrag{t3}[][]{{$t=3$ ms}}
\rotatebox{0}{\resizebox{!}{10cm}{\includegraphics{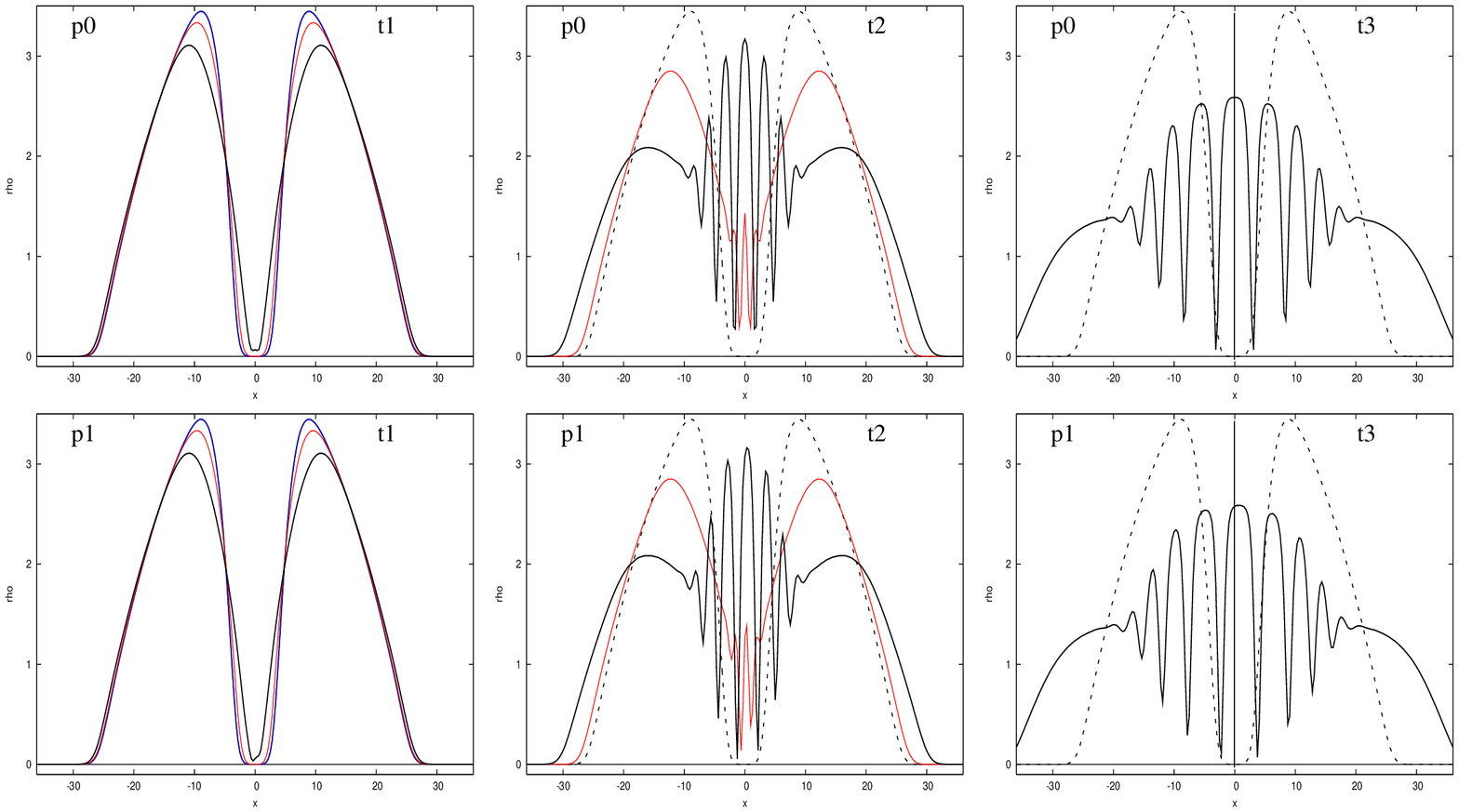}}}
\caption{\label{gp1}(Color online) Ballistic expansion dynamics of a fragmented double-well condensate ground state density approximated by the TDGP equation. The upper three panels display the dynamics of an initially stationary state, while the lower three panels display that of an initially $\pi/4$ phase offset between the left and right wells. Time increases from left to right, with snapshots taken every 0.25 ms so that a history of the dynamics is captured at 1 ms, 2 ms, and 3 ms after release of the trap at $t=0.$ Colors alternate to aid in visualization. Significant interference occurs, with almost 100$\%$ visibility, in this TDGP model, which is restricted to a single configuration for all time, and drastically contradicts the expansion dynamics of the TDMCBSCF fragmented ground state density and the results of the MIT interference experiment \cite{Andrews1997b} displayed in Fig. {\ref{frag1}}. Atomic correlations are thus seen to quench the mean-field effects for this early time dynamics.}
\end{figure*}

\subsubsection{Ballistic expansion dynamics within time-dependent Gross-Pitaevskii theory}
To further place the dynamics of these two condensate states into perspective, the single-orbital (and single-configurational) time-dependent Gross-Pitaevskii (TDGP) equation \cite{Gross1961a,Pitaevskii1961a} 
\begin{equation}
\begin{split}
i\hbar\dot\phi({\bf x},t)&=h({\bf x})\phi({\bf x},t)\\
&\ \ +(N-1)[{\textstyle\int}\phi^*({\bf x}',t)V({\bf x},{\bf x}')\phi({\bf x}',t)d^3x']\phi({\bf x},t)
\end{split}
\end{equation}
is integrated, at both zero and $\pi/4$ initial phase offsets, to mimic the $t=0$ density distribution of the fragmented ground state with $N=100.$ Note that this model is not an appropriate approximation; it is impossible, within the TDGP formalism to accurately describe a fragmented state, since, by definition, a fragmented state has macroscopic occupation in two or more Fock states \cite{Penrose56,Lowdin55a,Leggett2001a}. Since the TDGP dynamics involves only a single configuration, at all times, it would not be surprising for differences, perhaps even substantial differences, to arise between its dynamics and the dynamics associated with the TDMCBSCF theory, where even an initially single-configurational state may, in time, evolve into a complicated superposition of many configurations. Fig. \ref{gp1} presents the expansion dynamics of the TDGP approximation to the fragmented ground state density after release of the trapping potential. The upper three panels display the dynamics of an initially stationary state density, while the lower three panels display that of a TDGP density which has an initial phase offset of $\pi/4$ between the left and right halves of the TDGP orbital $\phi.$ Snapshots are taken, from left to right, every 0.25 ms so that the dynamics is captured at 1 ms, 2 ms, and 3 ms following the trap's release. For clarity, the histories ending at 2 ms and 3 ms show only a few time updates to the density, where, at 3 ms, the initial state is displayed in the background. In each panel, the density at the final integration step is plotted in black. Significant interference occurs, with near 100$\%$ visibility, within this single-configurational approximation: a result that was demonstrated in the linear regime, over a decade ago, in Ref. \cite{Naraschewski1996a}. This severe interference is in disagreement with the full TDMCBSCF theory presented in Fig. \ref{frag1}, where, as in the MIT interference experiment \cite{Andrews1997b}, data is taken in the high density regime where neither nonlinearity nor correlations can be neglected and interference fringes are subsequently observed with much less visibility. Here we see that atomic correlations almost quench the purely mean-field effects seen in TDGP theory. It is our opinion that this is a previously unappreciated result.

\section{Conclusion}
We have presented a first principles time-dependent quantum many-body theory of identical bosons to describe the many-body dynamics of a double-well BEC at zero temperature. Within a restricted two-state Fock space, our TDMCBSCF theory, which is derived from the TDVP, includes the full effects of the condensate's mean field and that of atomic correlation between atoms in different configurations, and, additionally, enjoys the complete and self-consistent coupling between the expansion coefficients of each configuration and the underlying mean-field orbitals. The TDMCBSCF evolution equations form a well-defined initial value problem and are an approximation to the exact time-dependent many-body Schr\"odinger equation. They have been implemented in an efficient and general numerical algorithm. In order to study the role of initial atomic correlation and mean-field effects upon the dynamics, we explore the ballistic expansion of an initially fragmented condensate ground state and an initially macroscopic quantum superposition state (also called a Schr\"odinger cat state) of the double-well BEC, with two different initial phase offsets between the left and right wells, following release of the external trapping potential. Brilliant matter-wave interference is observed at all times in the expansion dynamics of both the fragmented ground state and the superposition state, which is analogous to the double slit with all $N$ atoms interfering simultaneously. Remarkable sensitivity to small phase offsets between the left and right halves of the macroscopic quantum superposition are manifested in observable shifts in their interference patterns. It is shown that naively approximating the dynamics of the fragmented state with that of the TDGP approach leads to drastically different results than what is obtained in the full many-body theory and what is observed in experiment.

We note that, of course, decoherence will need to be taken into account \cite{zurek03}, especially for high-lying macroscopic quantum superposition states \cite{Moore06a}. In addition, if observations are being made then quantum back-action should also be included \cite{Dalvit02,Percival98}. However, getting the many-body quantum physics correct in the absence of these effects is a prerequisite to the proper treatment of decoherence and observation. The former is the contribution of this paper.

While in the final preparation of this paper, an article of similar intent has appeared on the arXiv; see Ref. \cite{Ced06b}. However, the short length of Ref. \cite{Ced06b} has prevented the presentation of any real details and, consequently, it is not possible to make a detailed comparison to the present work. Furthermore, we note that the numerical examples presented here are quite different from the fragmentation example in Ref. \cite{Ced06b}.

The authors gratefully acknowledge financial support from the National Science Foundation through the grant PHY 0140091. Both D.J.M. and W.P.R. have benefited from conversations with Dr. Hans-Dieter Meyer of the University of Heidelberg and Dr. Erik Deumens of the University of Florida regarding the equivalence of different time-dependent variational principles in quantum mechanics. Furthermore, D.J.M. and W.P.R. wish to thank the anonymous referee who pointed out a valuable concern with an original version of our manuscript that led us to rethink and clarify the representation of the initial TDMCBSCF state vector used in Secs. IV. B. 1 and 2.

\bibliography{jila,thesis,mal,ref}

\begin{thebibliography}{57}
\bibitem[{fn8()}]{fn8}
\bibinfo{note}{For notational simplicity, the explicit time dependence of the
  dynamical variables may be omitted when it is reasonable to do so.}
\end{thebibliography}
\end{document}